\documentclass{elsarticle}  

\usepackage{amssymb}

\journal{Planetary and Space Science}

\begin{document}

\begin{frontmatter}



\title{Europa's structural conditions for the existence of subsurface ocean and the absence of metallic core--driven magnetic field}


\author{Jun Kimura}

\affiliation{organization={Department of Earth and Space Science, Osaka University},
            addressline={1-1 Machikaneyama-cho}, 
            city={Toyonaka},
            postcode={560-0043}, 
            state={},
            country={JAPAN}}
\begin{abstract}
During the Galileo spacecraft's flyby of Europa, magnetic field measurements detected an inductive signal due to the response of Europa's interior conductors to temporal fluctuations in the Jovian magnetic field.
In contrast, no signatures of intrinsic magnetic field originating from the dynamo motion in the metallic core were acquired.
These measurements suggest that a global sub--surface ocean containing electrolytes exists beneath the solid ice shell and that the metallic core lacks convection.
Europa's interior is expected to be divided into the metallic core, rocky mantle and hydrosphere based on the moment of inertia factor estimated from gravity field measurements.
Specifically, the thickness of the outermost water layer is 120\,--\,170 km, and the radius of the metallic core is 0.12\,--\,0.43 times the surface radius.
No systematic investigation of Europa's internal evolution has been conducted to estimate the current state of the subsurface ocean and to explain the absence of a core dynamo field within such uncertainty for internal structure and material properties (especially ice properties).
Herein, I performed a numerical simulation of the long--term thermal evolution of Europa's interior and investigated the temporal changes in the ocean thickness as well as the temperature and heat flow of the metallic core.
If the ice reference viscosity is greater than 5$\times$10$^{14}$\,Pa\,s, the sub--surface ocean can persist even in the absence of tidal heating.
In the case of a tidal heating of 10\,mW/m$^{2}$ and 20\,mW/m$^{2}$, the ice shell thickness is $\le$\,90\,km if the ice reference viscosity is $\ge$\,1$\times$10$^{15}$ and 1$\times$10$^{14}$\,Pa\,s, respectively.
Regardless of the ice reference viscosity, if the tidal heating is $\ge$\,50\,mW/m$^{2}$, the shell thickness will be $\le$\,40\,km.
The thermal history of the metallic core is determined by the hydrosphere thickness and the metallic core density, and is unaffected by variations in the ice shell (ocean) thickness.
Preferred conditions for the absence of the core dynamo include CI chondritic abundance for the long--lived radioactive isotopes, lower initial core--mantle boundary (CMB) temperature and thicker hydrosphere.
The core may be molten without convection if the composition is near the eutectic in a Fe--FeS alloy, or not molten (without convection) if the composition is near the Fe or FeS endmember.
Specifically, if the rocky mantle has a CI chondritic radioisotope abundance, any core composition and hydrosphere thickness allow the absence of the core dynamo if the initial temperature at the CMB is lower than 1,250 K.
If the rocky mantle has the ordinary chondritic radioisotope abundance, or a higher initial temperature ($\sim$1,500 K) at the CMB, the core density lower than 6,000\,kg/m$^{3}$ is preferred for the absence of the core dynamo.
In the case of the core composition near the eutectic one, a hydrosphere thicker than 150\,km is required for the lacking core dynamo.
The lower pressure of Europa's rocky mantle due to its thinner hydrosphere compared with that of Ganymede may facilitate heat transfer in the mantle, lowering its temperature and making dynamo motion more challenging.
\end{abstract}


\begin{highlights}
\item A set of evolution models that satisfy the current Europa with an ocean and no core-driven magnetic field are presented.
\item Required conditions for the ice viscosity and the tidal heating rate to sustain the ocean at present was found.
\item Possible range of interior structure was narrowed down, compared to models based only on moment of inertia.
\end{highlights}

\begin{keyword}


Ocean \sep Moon \sep Thermal history \sep Interior \sep Magnetic field
\end{keyword}

\end{frontmatter}


\section{Introduction}
\label{sec1}

The Jovian moon Europa is a primary target for the search of extraterrestrial life because it is believed to have a salty ocean beneath its solid ice shell, as inferred from the detection of a magnetic field induced in a ocean during the Galileo spacecraft's closest approach to Europa \citep{kivelson00}. 
The only constraint currently available on the interior structure of Europa is the moment of inertia derived from gravitational measurements.
The hydrosphere encompassing the subsurface ocean and the superficial ice shell with a thickness of 200\,km or less must be separated from an inner rocky--iron component, based on measurements of Europa's gravitational field and resultant moment of inertia factor of 0.346\,$\pm$\,0.005 \citep{anderson98}. 
Although the thickness and depth of the ocean (i.e., ice shell thickness) are not precisely determined, geomorphological interpretation \citep[e.g.,][]{pappalardo99,schenk02} suggests that it exists several tens of kilometres beneath the surface.
In contrast, the dipole magnetic field originating from the dynamo activities in Europa's metallic core, such as that found on the outer moon, Ganymede, has not been confirmed.
Analysis from the moment of inertia factor suggests that Europa's deep interior is differentiated into a metallic core and an overlying rocky mantle assuming that Europa underwent a hot thermal evolutions \citep{anderson98}.
Recent numerical model assumed that Europa accreted at low temperatures (e.g,$\sim$200 to 300\,K) suggests that the metallic core formation may be delayed billions of years after accretion \citep{trinh23}.
It may support a recent value for the moment of inertia factor, 0.3547\,$\pm$\,0.0024 based on a re--investigation of the Galileo gravity data, suggesting that Europa may have thinner hydrosphere and lesser density of the rocky mantle, although the uncertainty in the gravitational flattening J$_{2}$ is large \citep{casajus21}.
Note that no conclusive evidence for the existence of a core on Europa has been obtained.
This current internal state is the consequence of a long--term evolution governed by a balance between internal heating and cooling, which is controlled by the volume of each layer.
For example, differences in the hydrosphere thickness permitted by the moment of inertia exceed 10\% of Europa's total volume, and the estimated volume of Europa's metallic core varies by a factor of 50 or more.
This could have a significant impact on its thermal evolution; therefore, a comprehensive numerical simulation with a diverse range of interior structures is required to narrow down the possible evolution of Europa.
Additionally, ice III, a high-pressure (HP) ice phase, may form if the hydrosphere is sufficiently thick and cold in the deepest depths.
However, the effect of ice III on Europa's thermal evolution has not been investigated.

Another Jovian moon Ganymede has an intrinsic global magnetic field that originated from a core dynamo activity that has not been confirmed on Europa.
Within the constraints imposed by the moment of inertia and bulk density, the overall size of Ganymede, excluding the outer hydrosphere, i.e., total size of the rocky mantle and the metallic core, is estimated to be approximately 1,534\,--\,1,634\,km \citep{anderson98,sohl02}, which is slightly larger but similar to that of Europa. 
Despite the similar proportions of rock and metallic components on both moons, the current state of the metallic core is substantially different.
It implies that even if Europa has a metallic core, it is either not molten or is molten but not convective at present.
It is observed that Europa's current interior is capable of sustaining a subsurface ocean but cannot develop a core dynamo.
Exploring the possibility of such a thermal state within the uncertainty of the interior structure is beneficial for constraining the actual interior structure and its long--term evolutionary history.

The stability of the sub--surface ocean and the driving force of the dynamo activity in the core is highly dependent on the thermal history and interior structure.
In a three--layered interior made of a hydrosphere, rocky mantle, and metallic core, the rocky mantle often contains a dominant heat source, which is the heat released from the decay of radioactive isotopes.
A body with a small and dense core overlain by a thick rocky mantle would have a larger heat budget than a body with a large and less dense core overlain by a thin rocky mantle.
Furthermore, either or both a smaller core and thicker hydrosphere would elevate mantle pressure, affecting the melting temperature and, consequently, the viscosity of the mantle.
All of these factors critically control the interior thermal evolution.

Generally, the initial temperature of icy bodies shortly after accretion is not considered to be particularly high, unlikely to reach the melting point of rock at around 1,600\,K for olivine \citep[e.g.][]{schubert86}.
In this case, the long--lived radioactive isotopes in the rock would heat up the interior after the end of accretion, and the quantity of rocks and heat transfer efficiency in that region would regulate the resultant thermal state.
\citet{kimura09} performed numerical calculations for the thermal history of Ganymede for various internal structures and investigated the structural conditions yielding a thermally--driven dynamo activity in the metallic core at present based on two conditions: the temperature at the core--mantle boundary (CMB) must exceed the melting point of a metallic core for an assumed composition, and the heat flux through the CMB must exceed the heat flux conducted along the core adiabat \citep{stevenson03,breuer15}. 
As a result, the inferred range of the interior structure can be narrowed down, compared with the range only based on the moment of inertia.

This study aims to constrain the conditions for Europa leading to the current existence of a subsurface ocean through an exploratory numerical investigation of the effects of different volumetric ratios of the layers, ice rheologies and tidal heating rates.
Another objective of this study is to confirm that the current thermal state of Europa's interior is compatible with a lack of dynamo activity, and to identify the structural condition that is incapable of driving the dynamo using numerical calculations.
Here, the simulation of Europa's thermal history to determine the admissible range of structures that are consistent with the current state are demonstrated.

First, the current understanding and uncertainties about Europa's interior structure are introduced using the moment of inertia as a reference.
The framework of the long--term thermal evolution model is then briefly described in Section\,2.
In Section 3, typical evolutionaly scenarios are presented, and then the potential depth and thickness of the ocean are determined based on the viscosity of the ice, tidal heating and structural uncertainties.
Additionally, the model can determine the thermal history of the metallic core and then constrain the core properties that lead to an absence of a core dynamo.
In Section 4, several discussions and future perspectives are provided.
Finally, the main conclusions are summarised in Section 5.
\section{Methods}
I performed numerical simulations of Europa's global thermal evolution under a wide range of parameter conditions to investigate the stability of the sub--surface ocean.
One of the major methods to calculate planetary thermal evolution is a parameterised convection model, which is a dimensionless energy balance model employing a scaling law between the Nusselt number (Nu) and the Rayleigh number (Ra).
This method has a drawback of making it unclear which parameter values are appropriate when calculating Ra because the thermophysical properties (e.g. the viscosity of the convective material) largely depends on temperature and depth. 
Although a 3D numerical scheme is another method that avoids this problem, their high computing costs limit the number of calculations that may be performed.
In this study, I employed a 1D method that is devoid of the aforementioned problems.
This method utilises the mixing length theory (MLT) to estimate the convective heat flux in the sub--solidus regime and has been applied to terrestrial planets and icy satellites in previous studies \citep{sasaki86,abe93,senshu02,kimura09,wagner11,kamata18}.
The MLT requires much lower calculation costs because this is a 1D scheme compared with the 3D modelling.
Consequently this scheme is suitable for parameter studies. 
And another advantage is that all the parameters can be determined locally, which can easily be applied to a case whose physical properties in the convective region vary significantly with depth.
This theory is consistent with the 3D calculations with the relative errors for the Nusselt number is less than 2\% for the isoviscous case and less than 10\% for the stagnant lid case \citep{kamata18}.
In the following sections, the interior model is described, followed by the solved equations below.
\subsection{Model for the interior structure}
The moment of inertia factor of 0.346\,$\pm$\,0.005 for Europa derived from gravitational measurements by the Galileo spacecraft suggests that Europa's interior is differentiated into a hydrosphere, rocky mantle and central metallic core \citep{anderson98,sohl02}.
Again, note that no definitive evidence for the existence of the metallic core has been acquired, and Europa would start from an incompletely differentiated state \citep[e.g.][]{travis12,bierson20,trinh23}

\begin{figure}
\centering
\includegraphics[scale=0.25]{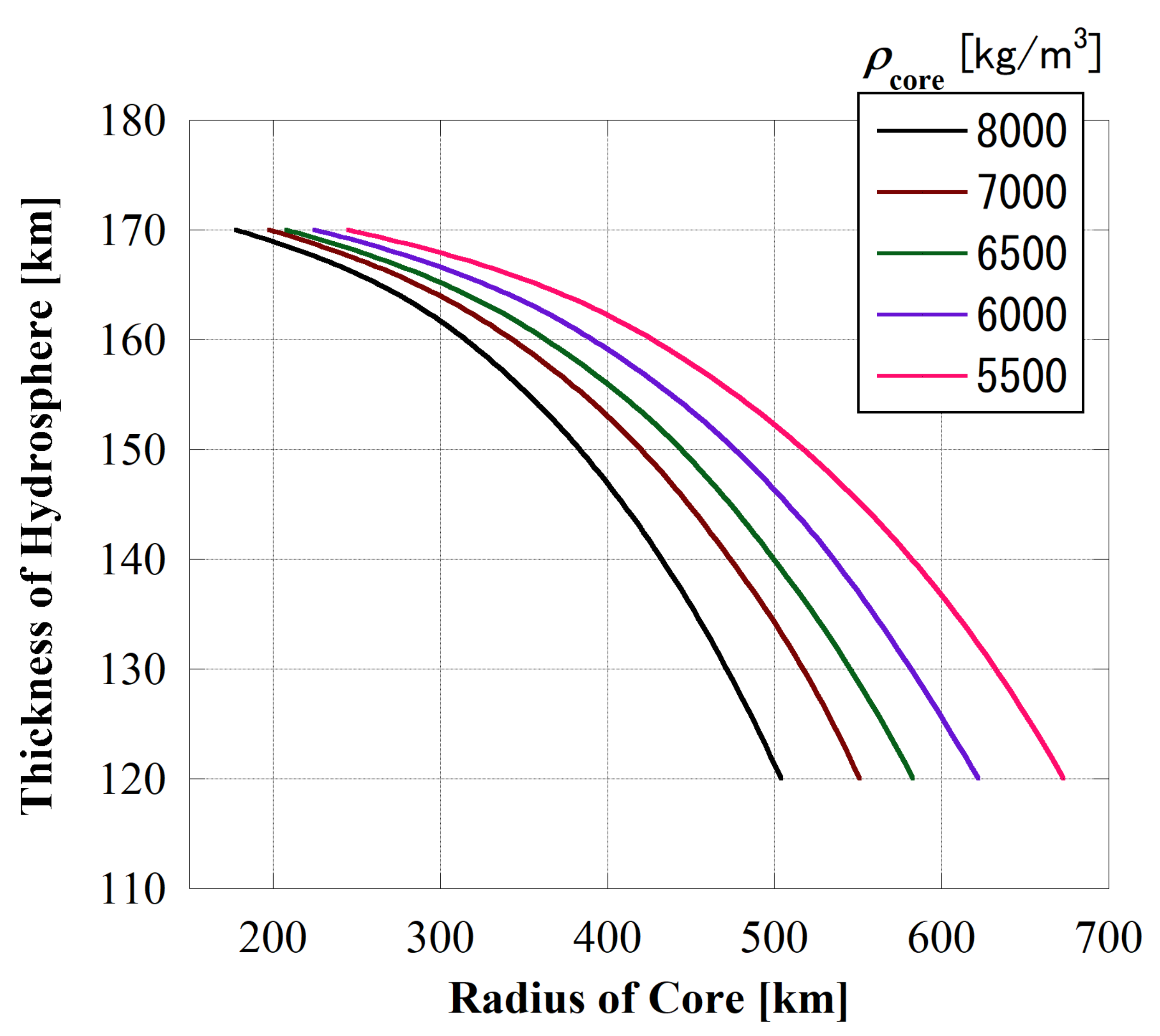}
\caption{Possible range of Europa's metallic core radius versus the thickness of the hydrosphere based on the constant--density shell models for the Europa's interior structure with an average density of 1,050\,kg\,m$^{-3}$, depending on the core density from 5,500 to 8,000\,kg/m$^{3}$. Modified from Sohl et al. (2002).}
\label{fig:interior}
Fig. \ref{fig:interior}.
\end{figure}

Figure \ref{fig:interior} depicts the inferred range of simple constant--density shell models for the interior structure of Europa based on an estimate by \citet{sohl02}, representing the relationship between the metallic core radius and the thickness of the hydrosphere depending on the metallic core density.
The reference surface radius is assumed to be 1,565\,km.
The density of the hydrosphere is assumed to be 1,050\,kg/m$^{3}$ as an averaged value for the possible existence of HP-ice phase at the bottom of the hydrosphere \citep{anderson98,sohl02}.
If the pressure in the hydrosphere exceeds about 207\,MPa and the temperature is below the melting point of 251\,K, ice III appears in the bottom of the hydrosphere.
In the structural model here, this corresponds to a hydrosphere thickness exceeding about 165\,km.
Once the HP--ice appears at the seafloor, it negatively affects the stability of the sub--surface ocean because the heat from the rocky mantle can no longer directly heat the sub--surface ocean.
The regime represented by solid lines satisfies the observed value of Europa's bulk density and the moment of inertia factor.
This demonstrates that the inferred structure has a high degree of uncertainty.
The core radius ranges between approximately 180\,km and 680\,km.
It is mostly a result of the uncertainty in the core density, which corresponds to the admissible sulfur content range between 0 and 36.5 wt\% in the Fe--FeS system. 
For example, the thickness of the hydrosphere, $D_{H_{2}O}$, at 120\,km and 170\,km accounts for 21.5\% and 29.2\% of Europa's total volume, respectively, and variation of the metallic core radius between 180\,km and 680\,km changes its volume by approximately 54 times.
Such a considerable difference in the volume of each layer can greatly influence the thermal history; hence, a comprehensive numerical analysis for an inferred wide range of interior structure is useful for determining the actual evolutional history, specifically to explore the current depth, thickness and time variations of the sub--surface ocean, and to determine the temperature and heat flux at the CMB that prevent a dynamo in the core.

The inferred range of the interior shown in Figure \ref{fig:interior} is consistent with the previous estimates \citep{anderson98,sohl02}, and in particular, if the metallic core radius is small (for example, below about 400\,km for a core density of 8000\,kg/m$^{3}$ and below about 550\,km for a core density of 5,500\,kg/m$^{3}$), the density of the rocky mantle exceeds the bulk density of Io, 3,526\,kg/m$^{3}$. 
Note that such high degrees of metal enrichment are unlikely for a body accreted farther in the circum-Jovian nebula than Io \citep{anderson98}. 
Nevertheless, it is important to investigate the different volumetric ratio between the core and the mantle on the thermal evolution, given the lack of definitive measurements for the interior structure of Europa.
The amount of sulfur in the core is not known and depends strongly on the primordial bulk composition and the conditions during metal--rock separation.
Estimates for sulfur amount predict both a composition more or less sulfur--rich than the eutectic for various types of chondrites.
For bulk sulfur contents of 1--2\,wt\% in ordinary chondrites and 3--6\,wt\% in carbonaceous chondrites, the metallic core that formed from chondritic precursors will contain 22--36\,wt\% and 6--21\,wt\% sulfur, respectively \citep[e.g.][]{kuskov01,daswani21,bercovici22}.
Thus, the S\,=\,22\,wt\% in this study corresponds to the lowest value for the carbonaceous chondrite.

Since the model here assumes that segregation does not occur in the core (no inner core growth), radius of the metallic core does not change with time.
In addition, the rocky mantle is assumed to did not go through a phase of hydration and does not go through a phase of dehydration.
Thus, the thickness of the rocky mantle is fixed throughout each calculation.
If Europa formed at much colder environment, less than about 550\,K for accretional temperature, Europa might end accretion as a mixture of hydrous silicates and metal \citep{trinh23}.
In this case, when the temperature reaches the dehydration point ($\sim$550 to 900\,K) due to the radiogenic heating and the tidal heating, the temperature increase would be buffered for several hundred million years by endothermic reactions associated with dehydration.
Thereafter, the viscosity would increase significantly as dry olivine--like composition and undergo a further temperature increase.
The maximum temperature is expected to be 100--200 K lower, making it difficult to satisfy the melting conditions for the metallic core.
Including the possibility that the metallic formation might not be achieved, such colder formation scenario may also be a reason in the absence of a core--driven magnetic field on Europa.

A recent re--investigation of the Galileo gravity data yielded a higher value for the moment of inertia of 0.3547\,$\pm$\,0.0024 suggesting that Europa may have a thinner hydrosphere and a less dense interior \citep{casajus21}.
This effect to my work will be discussed in Section \ref{subsec:moi_03547}.

\subsection{Numerical model for thermal evolution}
\begin{figure}
\centering
\includegraphics[scale=0.4]{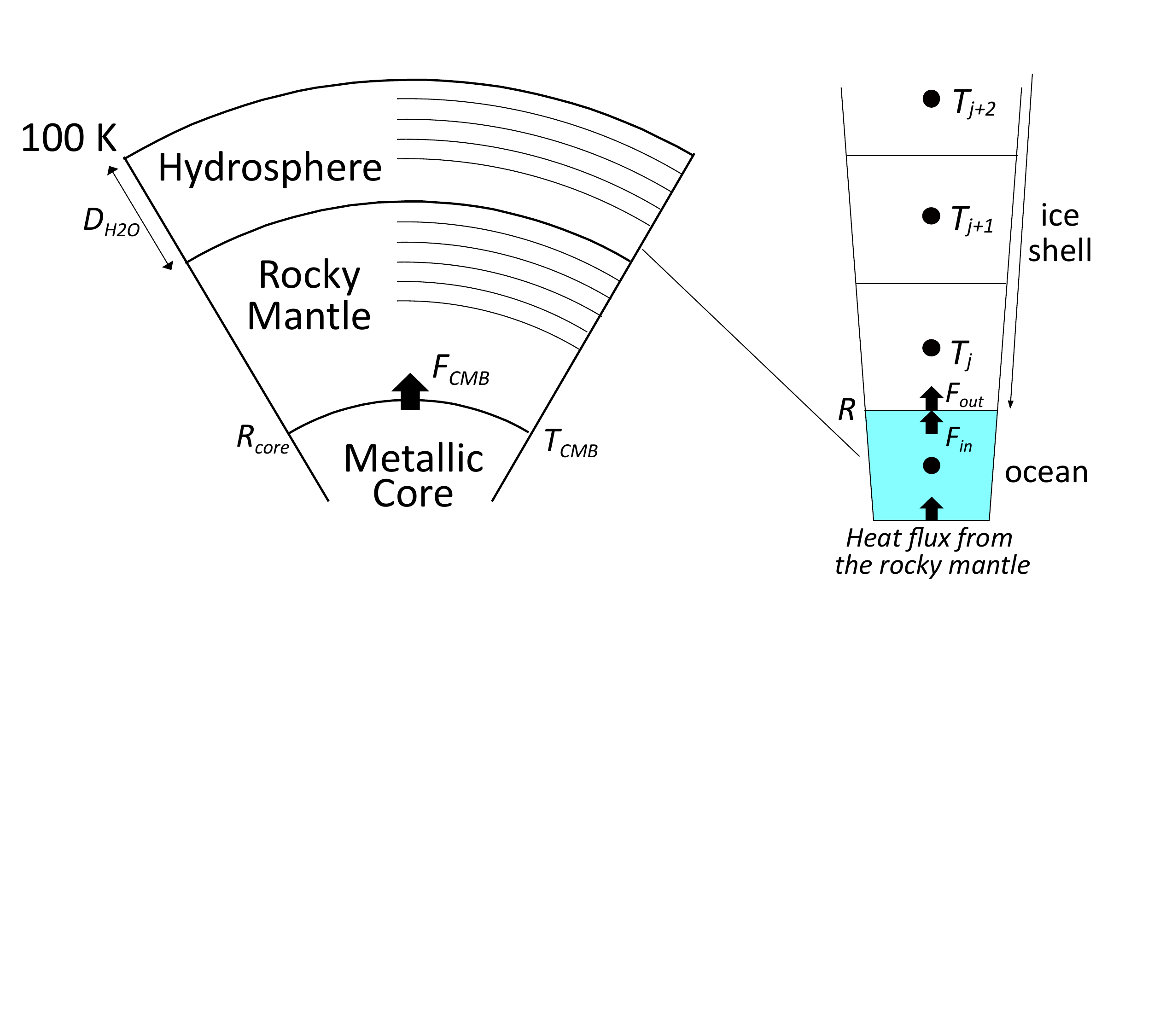}
\caption{Schematic diagram of the thermal evolution model setup.}
\label{fig:model_sketch}
Fig. \ref{fig:model_sketch}.
\end{figure}

It is considered that convection and conduction are responsible for heat transfer in Europa, while surface radiative heat transfer is neglected. 
The one--dimensional heat transfer equation from the CMB to the surface is solved.
The model setup is described in Figure \ref{fig:model_sketch}.
The following is the general equation for heat transfer:
\begin{equation}
\rho C_{p}\frac{dT}{dt}=-\frac{1}{r^{2}}\frac{d}{dr}\left(r^{2}F_{cond}+r^{2}F_{conv}\right)+\rho Q
\label{eq:heat_transfer}
\end{equation}
where $\rho$ is density, $C_{p}$ is specific heat, $T$ is temperature, $t$ is time, $r$ is radial distance from the centre, and $Q$ is heat production rate per unit mass.
The conductive and convective heat fluxes are given by
\begin{equation}
F_{cond}=-k\frac{dT}{dr},
\end{equation}
\begin{equation}
F_{conv}=-k_{v}\left\{\frac{dT}{dr}-\left(\frac{dT}{dr}\right)_{ad}\right\},
\end{equation}
where $k$ is thermal conductivity, and $k_{v}$ is {\it effective} thermal conductivity, which emulates the effect of thermal convection.
$k_{v}$ is expressed as follows \citep[e.g.,][]{abe93,kimura2020}:
\begin{equation}
\!k_{v}\!=\!\left\{
\begin{array}{ll}
\displaystyle{-\frac{\rho^{2}C_{p}\alpha g\ell^{4}}{18\eta}
\left\{\frac{\partial T}{\partial r}
\!-\!\left(\frac{\partial T}{\partial r}\right)_{ad}\right\}}
& \textrm{for }\;\displaystyle{\frac{\partial T}{\partial r}
\!<\!\left(\frac{\partial T}{\partial r}\right)_{ad}} \\
0 & \textrm{for }\;\displaystyle{\frac{\partial T}{\partial r}
\!>\!\left(\frac{\partial T}{\partial r}\right)_{ad}}
\end{array}
\right.
\end{equation}
where $g$ is the gravitational acceleration, $\alpha$ is thermal expansion coefficient, and $\eta$ is viscosity.
($dT/dr)_{ad}$ is the adiabatic temperature gradient given by
\begin{equation}
\left(\frac{dT}{dr}\right)_{ad}=-\frac{\alpha gT}{C_{p}}.
\end{equation}
Note that the both $dT/dr$ and $(dT/dr)_{ad}$ are negative; a positive convective heat flux appears only if the temperature gradient is steeper than the adiabatic temperature gradient.
$\ell$ is the mixing length, a parameter governing the convective heat flux that is a function of the convective region's depth.
The detailed deviation and explanation for the mixing length are described in \citet{kamata18}.
$\ell$ linearly increases with depth to the peak value $b$ until it reaches the peak depth $a$, and then it decreases linearly with depth as follows:
\begin{equation}
\!\ell\!=\!\left\{
\begin{array}{ll}
\displaystyle{\frac{b}{a}(R_{top}-r)}
& \textrm{for }\;\displaystyle{r\!\ge\! R_{top}-aD} \\
\displaystyle{\frac{b}{1-a}(r-R_{bot})}
& \textrm{for }\;\displaystyle{r\!\le\! R_{top}-aD}
\end{array}
\right.
\end{equation}
where $R_{bot}$ and $R_{top}$ are the radii at the bottom and top of the layer, respectively; $D$ is the thickness of the layer.
The conventional MLT scheme uses $a=0.5$ and $b=0.5$ \citep[e.g.][]{senshu02,kamata18}, where $\ell$ is assumed to represent the distance to the nearest boundary of the layer to reproduce a $Nu\sim Ra^{1/3}$ relationship \citep[e.g.][]{sasaki86,abe93,senshu02,kimura09}.
For the case where a layer has a large curvature, that is, $f=R_{bot}/R_{top}>0.5$, the following modifying values of $a$ and $b$ are well consistent with the predictions obtained by the 3D calculations \citep{kamata18}:

\begin{equation}
a(f,\gamma)=a_{2}(\gamma)f^{2}+a_{1}(\gamma)f+a_{0}(\gamma),
\end{equation}
\begin{equation}
b(f,\gamma)=b_{2}(\gamma)f^{2}+b_{1}(\gamma)f+b_{0}(\gamma),
\end{equation}
\begin{equation}
a_{2}(\gamma)=-41.2\exp(-0.297\gamma)-0.456,
\end{equation}
\begin{equation}
a_{1}(\gamma)=58.6\exp(-0.292\gamma)+0.704,
\end{equation}
\begin{equation}
a_{0}(\gamma)=-21.0\exp(-0.290\gamma)+0.624,
\end{equation}
\begin{equation}
b_{2}(\gamma)=3.96\exp(-0.167\gamma),
\end{equation}
\begin{equation}
b_{1}(\gamma)=-6.93\exp(-0.178\gamma),
\end{equation}
\begin{equation}
b_{0}(\gamma)=2.90\exp(-0.127\gamma),
\end{equation}
\begin{equation}
\gamma=\frac{2c^{2}_{0}\Delta T}{2c_{0}T_{b}+c_{1}-\sqrt{c^{2}_{1}+4c_{0}c_{1}T_{b}}},
\end{equation}
where $\Delta T$ is the temperature difference across the layer, and $T_{b}$ is the temperature at the base of the layer.
$c_{0}=1.23/f^{1.5}$ and $c_{1}=E_{a}/R_{g}$ where $E_{a}$ is activation energy and $R_{g}$ is the gas constant.
In the range of the possible interior structure considered here (Fig.\ref{fig:interior}), the hydrosphere and the rocky mantle has an $f$ value of 0.89--0.92 and 0.12--0.49, respectively.
Thus, the modified value of $a$ and $b$ described above are employed for the solid ice layers (the ice shell and the HP-ice layer) and the conventional value ($a=b=0.5$) for the rocky mantle.
The heat transfer equation (\ref{eq:heat_transfer}) is solved using a finite difference method based on the control volume method \citep{patankar80}.
As the number of grids is kept constant, the grid size is recalculated to account for changes in the ice layer thickness.

A parameterised convection theory without internal heat sources is adopted for the sub--surface ocean.
Table \ref{tab:physical_properties} lists the material properties that were adopted.
\begin{table}
\caption{Physical and material properties adopted in the thermal evolution calculations.}
\label{tab:physical_properties}
\vspace{0.1cm}
\begin{center}
{\scriptsize
\begin{tabular}{lll} \hline 
Parameter & Symbol & Dimensional value\\ \hline 
Density of the liquid H$_{2}$O layer & $\rho_{liq}$
& 1,000\,kg/m$^{3}$\\
Specific heat of the liquid H$_{2}$O layer$^{a}$ & $C_{p,liq}$
& 4.2\,$\times$\,10$^{3}$ J/K\,kg\\
Thermal conductivity of the liquid H$_{2}$O layer$^{b}$ & $k_{c,liq}$
& 0.566\,W/m\,K\\
Thermal expansion coefficient of the liquid H$_{2}$O layer$^{b}$ & $\alpha_{liq}$
& 2.1\,$\times$\,10$^{-4}$/K\\
Density of the rocky mantle$^{c}$ & $\rho_{m}$
& 3,300\,--\,3,800 kg/m$^{3}$\\
Specific heat of the rocky mantle$^{a}$ & $C_{p,m}$
& 920\,J/K\,kg\\
Thermal conductivity of the rocky mantle$^{a}$ & $k_{c,m}$
& 3.0\,W/m\,K\\
Thermal expansion coefficient of the rocky mantle$^{a}$ & $\alpha_{m}$
& 2.4\,$\times$\,10$^{-5}$/K \\
Density of the metallic core$^{c}$ & $\rho_{core}$ & 5,500\,--\,8,000\,kg\,m$^{-3}$ \\
Specific heat of the metallic core$^{d}$ & $C_{p,c}$ & 800\,J\,K$^{-1}$kg$^{-1}$ \\
Thermal conductivity of the metallic core$^{e}$ & $k_{c,c}$
& 5.0\,W/m\,K\\
Thermal expansion coefficient of the metallic core$^{f}$ & $\alpha_{c}$
& 8.0\,$\times$\,10$^{-5}$/K \\ \hline
\end{tabular}
}
\end{center}
\vspace{.3cm}
${}^{a}$ \citet{kirk87}\\
${}^{b}$ \citet{hill62}\\
${}^{c}$ \citet{sohl02}\\
${}^{d}$ \citet{buffett96,desai86}\\
${}^{e}$ \citet{pommier18}\\
${}^{f}$ \citet{williams09}
\end{table}

The energy balance at the boundaries is evaluated to investigate the time evolution of the boundary position between the ice layer(s) and the ocean. 
The movement of these boundaries' positions is given by
\begin{equation}
\rho_{ice}L_{ice}\frac{dR}{dt}=F_{in}-F_{out},
\end{equation}
where  $\rho_{ice}$ is the density of ice, $L_{ice}$ is the latent heat of ice, $R$ is the position of the phase boundary, $F_{in}$ is incoming heat flux and $F_{out}$ is outgoing heat flux.

When an ocean is present, the temperatures at the bottom of the ice shell and at the top of the HP--ice layer are set to the melting point of ice $T_m$, which is given by
\begin{equation}
T_{m}=T_{m_{0}}+\frac{dT_{m}}{dP}P
\end{equation}
where $T_{m_{0}}$ is a constant, and $P$ is pressure (Tab.\ref{tab:properties_ice}).
When an ocean is not present, the temperature at the bottom of the ice shell is determined by equating the heat flux at the base of the ice shell with that at the top of the HP--ice layer or the rocky mantle, depending on the total thickness of the hydrosphere.
\begin{table}
\caption{Physical properties of H$_{2}$O ice polymorphs adopted in the thermal evolution calculations}
\label{tab:properties_ice}
\vspace{0.1cm}
\begin{center}
\begin{tabular}{lllll} \hline 
Parameter & Symbol & Unit & ice I & ice III \\ \hline 
Density$^{a}$ & $\rho_{ice}$ & kg/m$^{3}$ & 930 & 1165 \\
Latent heat$^{a}$ & $L_{ice}$ & kJ/kg & 284 & 235 \\
Melting temperature at P=0$^{b}$
 & $T_{m_{0}}$ & K & 273.2 & 243.6 \\
Slope of melting temperature$^{b}$
 & $dT_{m}/dP$ & 10$^{-7}$K/Pa & $-$1.063 & 0.3597 \\ \hline 
\end{tabular}
\end{center}
\vspace{.3cm}
${}^{a}$ \citet{hobbs74}\\
${}^{b}$ \citet{sotin98}\\
\label{aba:table3}
\end{table}

The hydrosphere is assumed to be pure H$_{2}$O and devoid of any contaminants.
The presence of contaminants in the hydrosphere, such as ammonia, can significantly reduce the melting temperature \citep[e.g.][]{lewis721,kimura2020}, leading to a thinner ice shell.
As the ice shell thickens due to a secular cooling, the concentration in the ocean increases, further depressing the melting point of ice. 
Such effect suppresses a growth rate of the ice shell.
Thickness of the ice shell at present results in a few km to few tens of km thinner compared to the pure water case according to an initial concentration.
The effects of the presence of ammonia, which has a large melting point depression about 90\,K for the concentration of 30\,wt\% in maximum, have been discussed in detail in previous studies \citep{grasset96,kimura2020}.
However, no indications of the presence of ammonia have yet been discovered on Europa and other Jovian moons.
Remote sensing of the surface of Europa with near-infrared instruments has suggested the presence of hydrated materials, including sulfate salts (e.g. MgSO$_{4}$). 
However, the melting point depression of such salts is only 2-3 K  even in eutectic composition (about 17\,wt\%) \citep{kargel98}, thus I neglected this effect in this work. 
NaCl that has been suggested on Europa's surface from recent telescopic spectroscopy \citep{trumbo19,trumbo22} has a larger effect for a melting point depression up to 21.1\,K for 23.3\,wt\% \citep[e.g.][]{lamas22}. 
Although such high concentration of NaCl has not been confirmed and calculations considering the presence of NaCl are out of scope of this study, the results of previous studies (e.g., Grasset and Sotin, 1996; Kimura and Kamata, 2020) considering the presence of ammonia suggest that the ice shell thickness at present becomes thinner by several to ten kilometers depending on the concentration.

In the ice shell, temperature dependencies of physical properties are considered.
The specific heat, thermal conductivity and thermal expansion coefficient of ice Ih are given by
\begin{equation}
C_{p,ice}=7.037\,T+185.0,
\end{equation}
\begin{equation}
k_{c,ice}=\frac{632.0}{T}+0.38-0.00197\, T,
\end{equation}
\begin{equation}
\alpha_{ice}=3.0\times\left(2.5\times 10^{-7}\,T-1.25\times 10^{-5}\right),
\end{equation}
respectively, where the unit of $T$ is Kelvin \citep{hobbs74, andersson05}.

The thermal conductivity for HP--ice (ice III) is expressed by \cite{andersson05} as follows:
\begin{equation}
k_{c,hp}=93.2\times T^{-0.822}.
\label{eq:condictivity_hp}
\end{equation}
Other physical property data are highly limited and have considerably higher uncertainties than those for ice Ih.
For simplicity, I used the values of ice Ih for HP--ice, except for the density and thermal conductivity.

The viscosity of ice strongly affects the efficiency of heat transfer in the ice shell.
The ice viscosity's large temperature dependency is well--approximated by
\begin{equation}
\eta_{ice}
=\eta_{ref,ice}\,\exp\!\left[\frac{E_{a}}{R_{g}}\left(\frac{1}{T}-\frac{1}{T_{m}}\right)\right]
\label{eq:viscosity}
\end{equation}
where $\eta_{ice}$ is the viscosity of ice, $\eta_{ref,ice}$ is the reference viscosity of ice and $E_{a} = 60$\,kJ\,mol$^{-1}$ \citep{goldsby01}.
A typical value of $\eta_{ref}$ is approximately 10$^{14}$ Pa\,s \citep[e.g.][]{hussmann15}, which is comparable with that of terrestrial glaciers, although it can vary largely depending on many parameters, such as grain size ($\eta_{ref,ice}$\,=\,10$^{14}$ Pa\,s is equivalent to a few tenths of a millimetre in grain diameter \citep{barr09}). 
Thus, in this study, $\eta_{ref,ice}$ is a free parameter ranging between $1.0 \times 10^{13}$ and $1.0 \times 10^{17}$ Pa\,s for ice Ih.
For the sake of simplicity and the absence of adequate data, I assumed the same equation and parameters for HP-ice.

It should be noted that convection in the ice shell (ice I) and in the HP-ice (ice III) layers is considered separately, even though these layers are in contact with each other because the Clapeyron curves for the endothermic phase change between ices I and III suggests two-layer convection rather than whole-layer convection \citep{bercovici86}.
It is hypothesised that as the HP--ice layer melted due to the heat from the rocky mantle, and the liquid water formed at the base of the HP-ice layer ascended instantaneously to the top of the HP-ice layer \citep{kalousova18,kimura2020}.

The physical properties of the rocky core, except for the viscosity, are assumed to be uniform (Tab. \ref{tab:physical_properties}). 
Here I modelled the rocky mantle as olivine mantle, which is a mineral often used to characterize a planetary mantle \citep{sohl02}, and considered the temperature--dependent core viscosity given by
\begin{equation}
\eta_{rock}
=\eta_{ref,rock}\,\exp\!\left[A\left(\frac{T_{m,rock}}{T}
\right)\right],
\label{eq:rock_viscosity}
\end{equation}
where $\eta_{ref,rock}$ is reference viscosity of the core, $A$ is a constant and $T_{m,rock}$ is the solidus temperature of the rocky mantle.
I adopted $\eta_{ref,rock}$\,=\,4.9\,$\times$\,10$^{8}$ Pa\,s, $A$\,=\,23.25 and $T_{m,rock}\,=\,1,600$\,K \citep{karato86,kimura09}.
Values of A have an uncertainty (or spread) among the experimental works, then another result for different values of A will be discussed in section \ref{subsec:rock_rheology}.

The decay of long-lived radioactive isotopes ${}^{238}$U, ${}^{235}$U, ${}^{232}$Th and ${}^{40}$K produces heat in the rocky mantle.
In this study, I considered the radioactive isotope abundances of CI chondrites \citep{lodders03} and ordinary chondrites \citep{Wasson535} (Tab.\,\ref{tab:radioelem}).
The concentration in the latter is approximately 30\% larger than the former, and as a result, the ocean thickness can be 2--4 times larger as discussed in Section \ref{subsec:abundances}.
The heat source term, $Q$, in equation (\ref{eq:heat_transfer}) for the rocky mantle is then described as follows:
\begin{equation}
Q=\sum_i c_{i}H_{i}\exp\left(-\lambda_{i}t\right)
\end{equation}
where $i$ represents isotopes, $c_{i}$ is initial concentration, $H_{i}$ is heat release, and $\lambda_{i}$ is the decay constant.
In this model, the radioisotope abundance is assumed to be the concentration rate according to the CI or ordinary chondritic composition and included in the energy equation as the heat source amount per unit mass.
A more detailed approach considering the feedback from the assumed chondritic composition, the rock mass fraction and the radioactive amount is outside the scope of the current work.
In addition, compressibility of each layer needs to be considered to make this model more realistic, which will be the subject of a future study.
\begin{table}
\caption{Concentration, energy production and decay constants of radioactive elements}
\label{tab:radioelem}
\vspace{0.1cm}
\begin{center}
\begin{tabular}{ccccc} \hline 
 & Concentration & Concentration & & \\
 & in CI& in Ordinary& & Half \\
 & Chondrites$^{a}$ & Chondrites$^{b}$ & Decay Energy$^{a}$ & Life$^{a}$ \\
 & $c_{i}$ & $c_{i}$ & $H_{i}$ & $\lambda_{i}$ \\
 & (ppb) & (ppb) & (10$^{-5}$ W/kg) & (Myr) \\ \hline 
$^{238}$U & $19.9$ & $26.2$ & $9.465$ & $4468$ \\
$^{235}$U & $5.4$ & $8.2$ & $56.87$ & $703.81$ \\
$^{232}$Th & $38.7$ & $53.8$ & $2.638$ & $14,030$ \\
$^{40}$K & $738$ & $1104$ & $2.917$ & $1277$ \\ \hline 
\end{tabular}
\end{center}
\vspace{.3cm}
$^{a}$ \citet{lodders03}\\
$^{b}$ \citet{Wasson535}
\end{table}

It is assumed that tidal dissipation is the only heat source for the ice shell and that its rate is time independent for the simplicity.
Although the tidal dissipation can actually occur at any depth, it is assumed that all the tidal dissipation occurs at the base of the ice shell because the heating rate is highest where the temperature is closest to the melting point \citep[e.g.][]{ojakangas89,tobie03,kamata18}. 
In this model, the tidal heating rate, denoted as $Q_{t}$, is explored between 0 and 100\,mW/m$^{2}$ which has been estimated by assuming a Maxwell rheology \citep{hussmann02,tobie03,sotin09}.

Tidal heating in the rocky mantle is neglected here.
If the rocky mantle is heated sufficiently to melt, even partially, through excitation of eccentricity and increased tidal dissipation through the orbital resonances with Io and Ganymede, the tidal dissipation becomes stronger, and the temperature increases further \citep{behounkova20}.
Consequently, this positive feedback may lead to thermal and melting runaways.
Although such a process is outside the scope of this paper, if it occurs before the differentiation between rock and metal, the core formation would be promoted.
If it occurs after the core formation, further heating of the rocky mantle would make the core more susceptible to a molten state and less likely to satisfy the cooling conditions (like as present Io does not have a dynamo field).
This could also cause for the lacking core--dynamo in the current Europa.

\subsection{Initial state and other settings}
As it is difficult to constrain the timing of differentiation or any episodes of tidal heating due to orbital resonances \citep{showman97}, all simulations begin with a completely differentiated structure at 4.5\,Ga \citep{kirk87}.
The core formation may occur at later stage if Europa formed at much colder temperature \citep{trinh23}.
Simple calculation of the energy released upon the differentiation results in a temperature increase of 100\,$\sim$\,200\,K, which is an order of magnitude smaller than the accretional energy \citep{schubert86}.
On the other hand, during the process of melting of metal, the latent heat is consumed, and thus temperature variations are buffered.
Although such processes during a separation between rock and iron components has not been investigated in detail, the energy consumption due to melting of metal and the energy release associated with rock-metal differentiation may not have much effect on the final state, since they act in opposing temperature changes. 
Rather, the fractionation and redistribution of radioactive isotopes in the rock during differentiation may affect the subsequent temperature state. This could be important for future work.

The initial hydrosphere is mostly molten (1\,km thickness ice shell in initial) and the primitive liquid water layer (primitive ocean) is overlying the rocky mantle and the metallic core.
The surface of the ice shell is fixed at $100$\,K throughout the calculation.
Temperature at the boundary between the primitive ocean and the rocky mantle is set to the melting temperature of the ice phase according to existing pressure conditions.
Even in the case where the initial hydrosphere is entirely frozen, the time evolution of the ocean thickness after it reaches its maximum at 1.0\,--\,1.5\,Gyr and the final ocean thickness are quite similar to the case when the initial hydrosphere is entirely molten \citep{kimura2020}.
Therefore, the initial thermal condition does not affect the long-term thermal evolution, specifically the evolution of ocean thickness.

The initial thermal structure in the rocky mantle is set to be a steady state with the melting temperature of the ice at the top boundary and the eutectic temperature in the Fe--FeS system ($1,250$\,K) at the bottom boundary (CMB) because the state immediately after the core formation is assumed to be the initial state here.
This initial temperature at the CMB can be regarded as a lower limit, whereas \citet{hauck06} proposed an initial temperature of $2,000$\,K there as the highest.
\citet{schubert86} provided the temperature increase of $\sim$\,1,300\,K for accretion and 100\,$\sim$\,200\,K for differentiation.
Even if the sub--Jovian nebula was balmier, the initial temperature may not have exceeded about $1,500$\,K \citep{kuramoto94}.
The selection of the initial temperature could effect the resultant thermal history.
Thus, additional settings with $1,500$\,K at the CMB is also investigated, and the differences in the result are detailed in Section \ref{sec:results}.

\subsection{Evaluating the dynamo activity in the metallic core}
Similar to prior research, this study employs a simple criterion for thermal convection in the metallic core to evaluate the operation of dynamo activity \citep{nimmo02,stevenson03,kimura09}.
The mathematical formulation of core thermal evolution is based on simple analytical models:
\begin{equation}
\frac{4}{3}\pi\rho_{core}C_{p,c}R^{3}_{core}\frac{dT_{CMB}}{dt}=-4\pi R^{2}_{core}F_{CMB}
\end{equation}
where $T_{CMB}$ is the temperature at the CMB. 
$\rho_{core}$, $C_{p,c}$ and $R_{core}$ are the density, the specific heat and the radius of the metallic core, respectively. 
$F_{CMB}$ is the heat flux through the CMB, which is calculated from the temperature gradient at the base of the mantle.
To assess the generation of thermal convection in the metallic core, I employed the following two simple conditions. 
The first condition is that the temperature at the CMB is higher than the melting point of the assumed core composition, implying that the metallic core is at least partially molten.
The second condition is that the heat flux at the CMB must also be greater than the adiabatic temperature gradient, $F_{ad,CMB}$\,=\,$k_{c}\alpha_{c} gT_{CMB}/C_{p,c}$, where $k_{c}$ is the thermal conductivity, $\alpha_{c}$ is the thermal expansivity and $C_{p,c}$ is the specific heat of the metallic core \citep{nimmo02,stevenson03}.
Values of material parameters for the core is listed in Table \ref{tab:physical_properties}.
Note that the thermal conductivity of 5.0\,Wm/K is adopted here, and this value dramatically changes with sulfur content; 53\,W/mK for S\,=\,0\,wt\%, 5.0\,W/mK for S\,=\,20.0\,wt\% and 3.8\,W/mK for S\,=\,36.5\,wt\% at the pressure of the Europa's core \citep{williams09}.
If the core has a larger conductivity, the adiabatic temperature gradient becomes large and the cooling condition would be difficult to satisfy.
Compositional convection may continue even if the heat flux of the core is less than $F_{ad,CMB}$, but it will cease if the heat extracted from the core drops to zero or less (i.e. the core starts heating).
Using these formulations and parameters, I conducted case study on thermal history simulations of various structures within the range depicted in Figure\,\ref{fig:interior}. 
If the results of a particular interior structure model satisfy both the current melting and cooling conditions, I considered this model to be a realistic approximation of Europa's interior.
\section{Results}
\label{sec:results}
I describe the results of my models and particularly the evolution of the subsurface ocean and the metallic core in Europa as a function of ice reference viscosities, core and mantle sizes, tidal heating rates and concentrations of radioisotopes.

\subsection{General trend for thermal history and ocean thickness}
\begin{figure}
\includegraphics[scale=0.28]{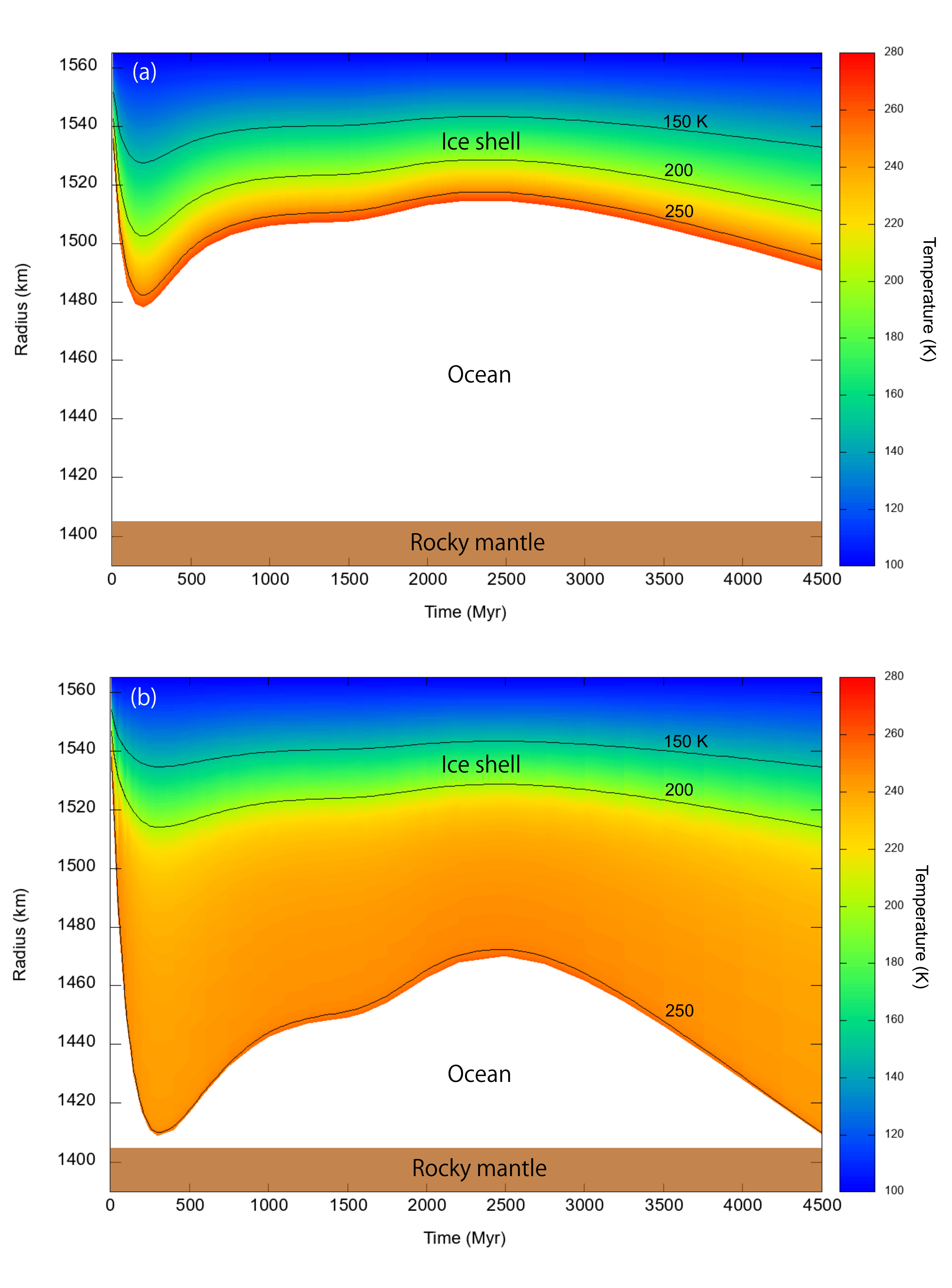}
\caption{Examples of result for the structure of the subsurface ocean and the ice shell of Europa as a function of radius and time. Assumed interior structure for the radius and density of the core is 364\,km and 6,500\,kg\,m$^{-3}$, respectively. The reference (melting point) viscosity of the ice ($\eta_{ref}$) are  1$\times$10$^{17}$\,Pa\,s (a) and 5$\times$10$^{14}$\,Pa\,s (b). For both models, CI chondritic abundances for long--lived radioactive isotopes are assumed in the rocky mantle.}
\label{fig:prof_s_CI}
Fig. \ref{fig:prof_s_CI}.
\end{figure}

\begin{figure}
\centering
\includegraphics[scale=0.6]{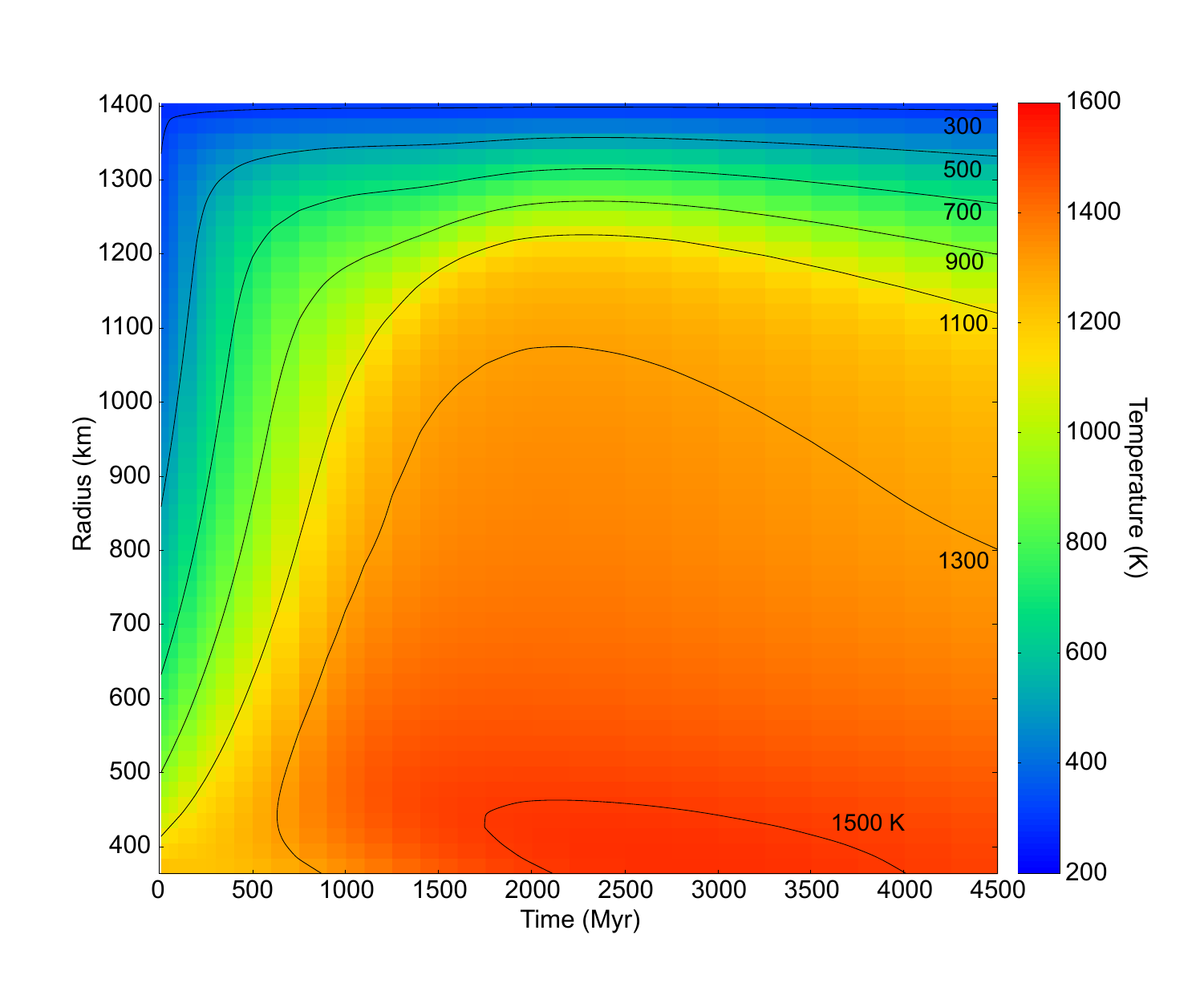}
\caption{Temperature profile changing with time of the rocky mantle for the same calculation setup as depicted in Fig.\,\ref{fig:prof_s_CI}a (the radius of the metallic core is 364\,km, and the entire thickness of the hydrosphere (D$_{H_{2}O}$) is 160\,km which corresponds to the upper boundary of the mantle of 1405\,km).
}
\label{fig:prof_m_CI}
\end{figure}

Figure\,\ref{fig:prof_s_CI} depicts the typical results for the time evolution of the ice shell and the subsurface ocean when the radius and the density of the metallic core are 364\,km and 6,500\,km/m$^{3}$, respectively.
Tidal heating is not included in this result.
The entire thickness of the hydrosphere (D$_{H_{2}O}$) is assumed to be 160\,km, and the reference (melting point) viscosity of the ice ($\eta_{ref}$) is 1$\times$10$^{17}$\,Pa\,s (Fig.\,\ref{fig:prof_s_CI}a) and 5$\times$10$^{14}$\,Pa\,s (Fig.\,\ref{fig:prof_s_CI}b). 
In the rocky mantle, tidal heating is not accounted for, and CI chondritic abundances for long--lived radioactive isotopes are assumed.
During its early stages, the ice shell rapidly grows, and the solidification process continues up to approximately 0.3\,Gyr in both cases for the ice reference viscosities, when the heat loss through the ice shell balances with the heat input from the rocky mantle.
The heat from the rocky mantle gradually increases due to the decay of radioactive isotopes, remelting the ice shell.
The rocky mantle is continuously heated by the decay energy of the radioactive isotopes up to 2.5\,Gyr (Fig.\,\ref{fig:prof_m_CI}), resulting in the thickening of the ocean.
The heat flux from the mantle is particularly increased by solid-state convection occurring in the rocky mantle from approximately 1.7\,Gyr, which causes the ice shell to shrink even further.
Although this convection maintains by approximately 3.0\,Gyr, after 2.5\,Gyr, the depletion of radioactive isotopes in the rocky mantle and secular cooling thickens the ice shell toward the present day.
Thereafter, as the heat of the rocky mantle depletes, secular cooling thickens the ice shell toward the present day.
In case of ice with a higher reference viscosity (Fig.\,\ref{fig:prof_s_CI}a), the ice shell is always conductive regardless of its thickness, whereas in the case of lower viscosity (Fig.\ref{fig:prof_s_CI}b), the ice shell is strongly convective, and the higher rate of heat transfer through the ice shell results in a thinner ocean.
The thermal evolution of the rocky mantle is not affected by the reference viscosity of ice.

\begin{figure}
\centering
\includegraphics[scale=0.6]{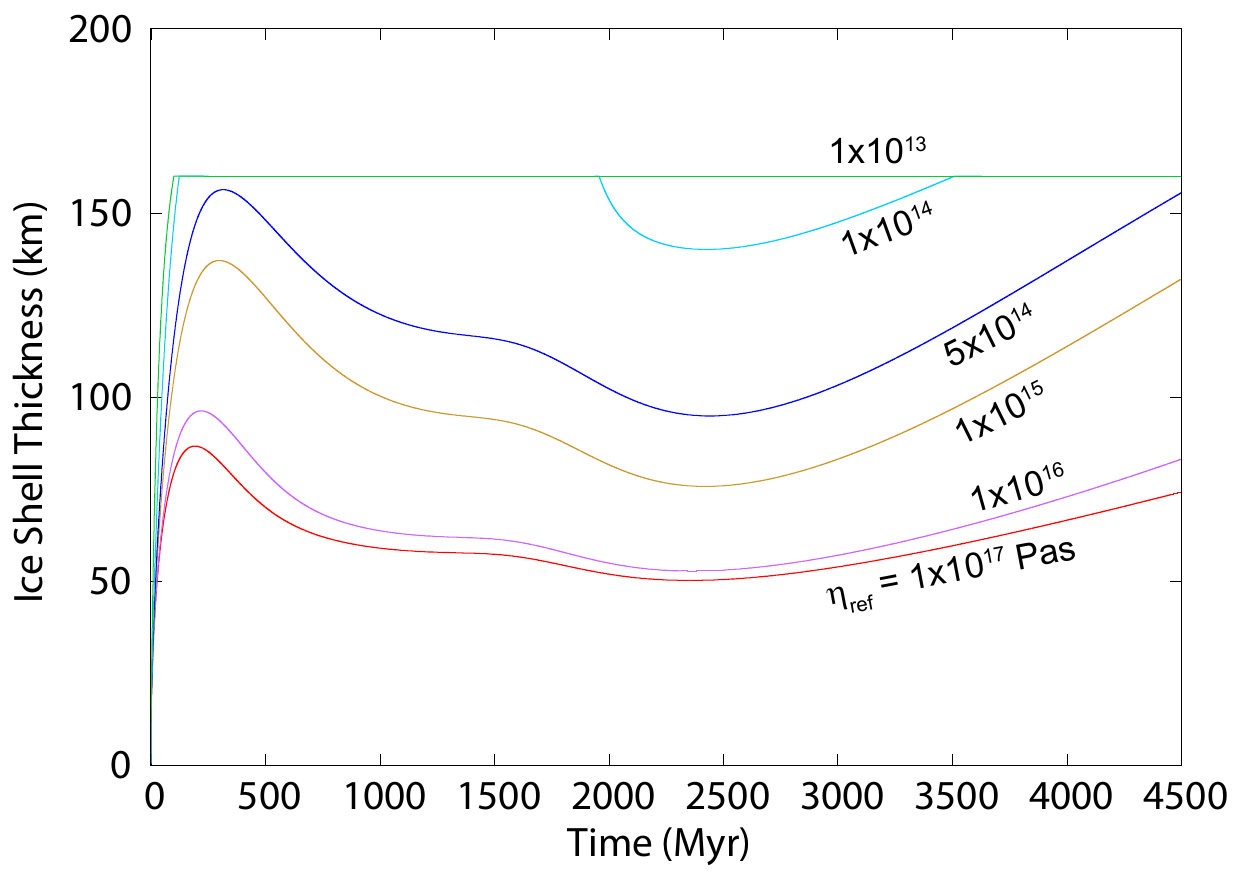}
\caption{Time evolution of the ice--shell thickness for different ice reference viscosities ($\eta_{ref}$) for the same calculation setup as depicted in Fig.\ref{fig:prof_s_CI}a ($\eta_{ref}$\,=\,1$\times$10$^{17}$\,Pa\,s) and Fig.\ref{fig:prof_s_CI}b (5$\times$10$^{14}$\,Pa\,s). A thickness of 160\,km indicates no ocean exists because D$_{H_{2}O}$\,=\,$160$\,km.}
\label{fig:prof_s_CI_etaref}
\end{figure}

Figure \ref{fig:prof_s_CI_etaref} demonstrates that the reference viscosity of ice predominantly determines the ice--shell thickness.
A greater viscosity results in a thinner ice shell (a thicker ocean).
In the case of lower viscosity, the ice shell becomes convective, and effective heat removal causes the ocean to freeze, which is consistent with previous works \citep[e.g.][]{kamata18,kamata19,kimura2020}.
While no ocean can form at all for $\eta_{ref}=$1$\times$10$^{13}$\,Pa\,s, it is temporally possible for the ocean to form between 2.0\,Gyr and 3.5\,Gyr in the case of $\eta_{ref}=$1$\times$10$^{14}$\,Pa\,s.
It should be noted that even in the absence of tidal heating, the subsurface ocean can be sustained throughout history if the reference viscosity is $\ge$ 5$\times$10$^{14}$\,Pa\,s, because the ice shell is conductive resulting in a lower heat transfer rate.

\begin{figure}
\centering
\includegraphics[scale=0.55]{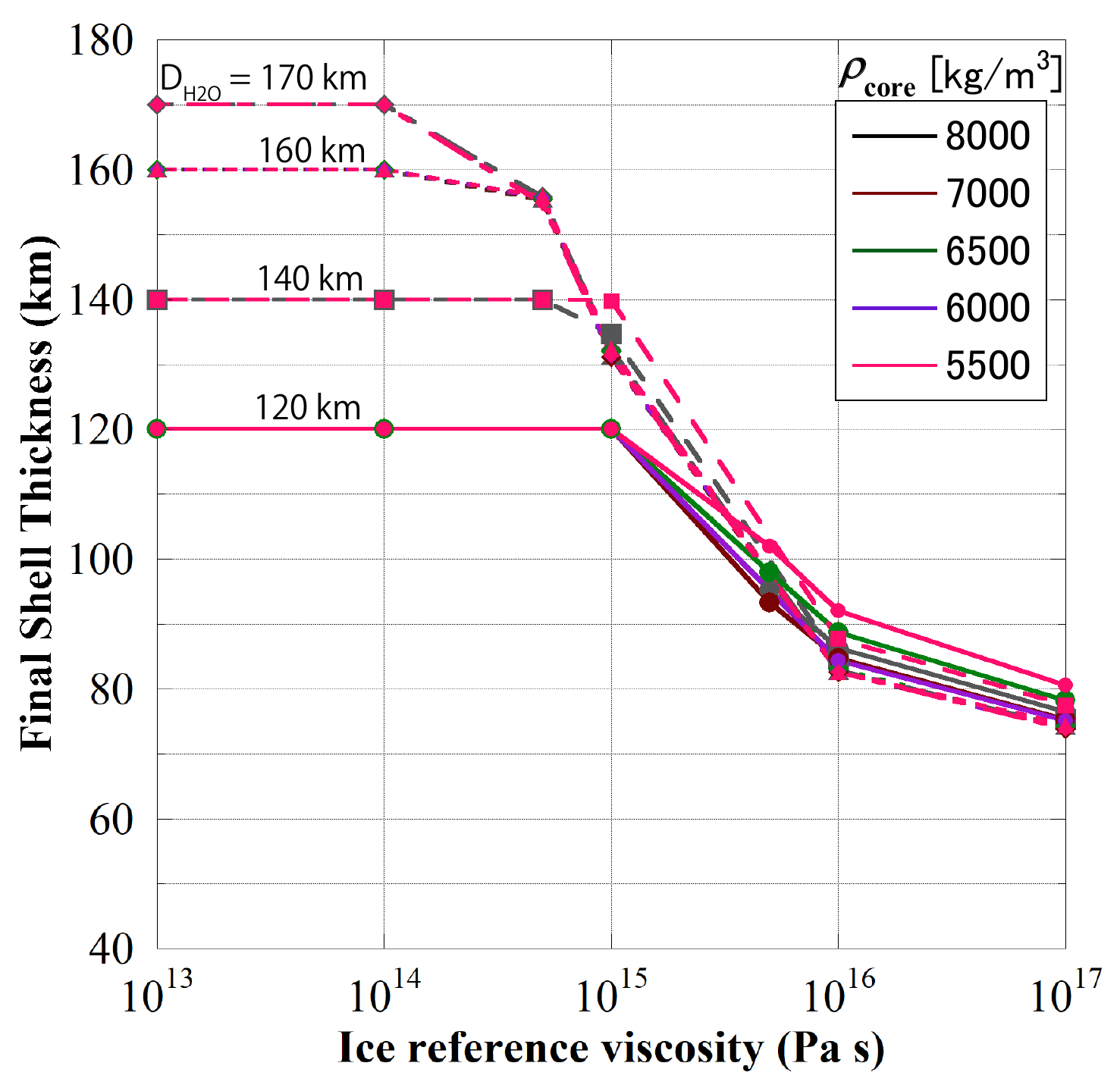}
\caption{Thickness of the ice shell (right) at present as a function of the ice reference viscosity under different structural conditions.
}
\label{fig:final_thickness_CI}
Fig. \ref{fig:final_thickness_CI}.
\end{figure}

Based on the admissible size range for each layer (i.e., the thickness of the hydrosphere, $D_{H_{2}O}$, and the core density, $\rho_{core}$) indicated in Figure\,\ref{fig:interior}, comprehensive calculations were performed on various interior structures.
Figure\,\ref{fig:final_thickness_CI} depicts the final thickness (i.e. at 4.5\,Gyr) of the ocean and ice shell under various computation settings. 
Generally, a larger hydrosphere thickness results in a thicker ocean.
Currently, the minimal reference viscosities of ice required to sustain the ocean at present (regardless of its thickness) are $1 \times 10^{14}$ Pa\,s for $D_{H_{2}O}=170$\,km and $1 \times 10^{15}$ Pa\,s for $D_{H_{2}O}=120$\,km.

In contrast, differences in the core density have little impact on ocean thickness, although differences in metallic core volume affect the amount of rocky mantle (amount of radioactive heat source). 
In the admissible range, for each value of $D_{H_{2}O}$, a 10\%\,--\,15\% difference in mantle mass only impacts the current ocean's thickness by $\le$\,10\,km.
In all theoretical scenarios, the final thickness of the ice shell is $\ge$\,70\,km.
There are numerous earlier studies that estimate the thickness of the ice shell.
The thickness estimates from a few kilometres to 50\,km are made using geological analyses of surface tectonic features and theoretical models for interior thermal condition \citep[e.g.][]{carr98,Rathbun98,pappalardo99,hoppa99,Greenberg00,hussmann02,nimmo03}.
According to impact crater studies, thicknesses range from 3 to 4\,km \citep{turtle01} to 19 km \citep{schenk02}.
In 3D numerical simulation, it suggests that the ice shell should be 50\,--\,90\,km thick \citep{vilella20}.
Calculating a steady-state internal heat balance considering a wide range of possible values and uncertainties for physical properties and layer thicknesses predicts the probability distribution of the ice--shell thickness to be 23-47\,km \citep{howell21}.

There is no admissible range that is consistent with previous estimates for the typical value of the reference ice viscosity $\eta_{ref}$\,=\,10$^{13}$\,--\,10$^{15}$\,Pa\,s, as all results for the ice--shell thickness are $\ge$\,90\,km for all structural parameters.
This means that another factor is needed to produce a thinner ice shell which is suitable for previous geological and numerical estimates.

In case of a higher initial temperature of T$_{CMB}$\,=\,1,500\,K, the final thickness of the ocean is very similar to the results in the case of T$_{CMB}$\,=\,1,250\,K.
Thus a different initial T$_{CMB}$ value has no effect on the final ocean thickness.

\subsection{Effect of tidal heating on ocean thickness}
\begin{figure*}
\centering
\includegraphics[scale=0.6]{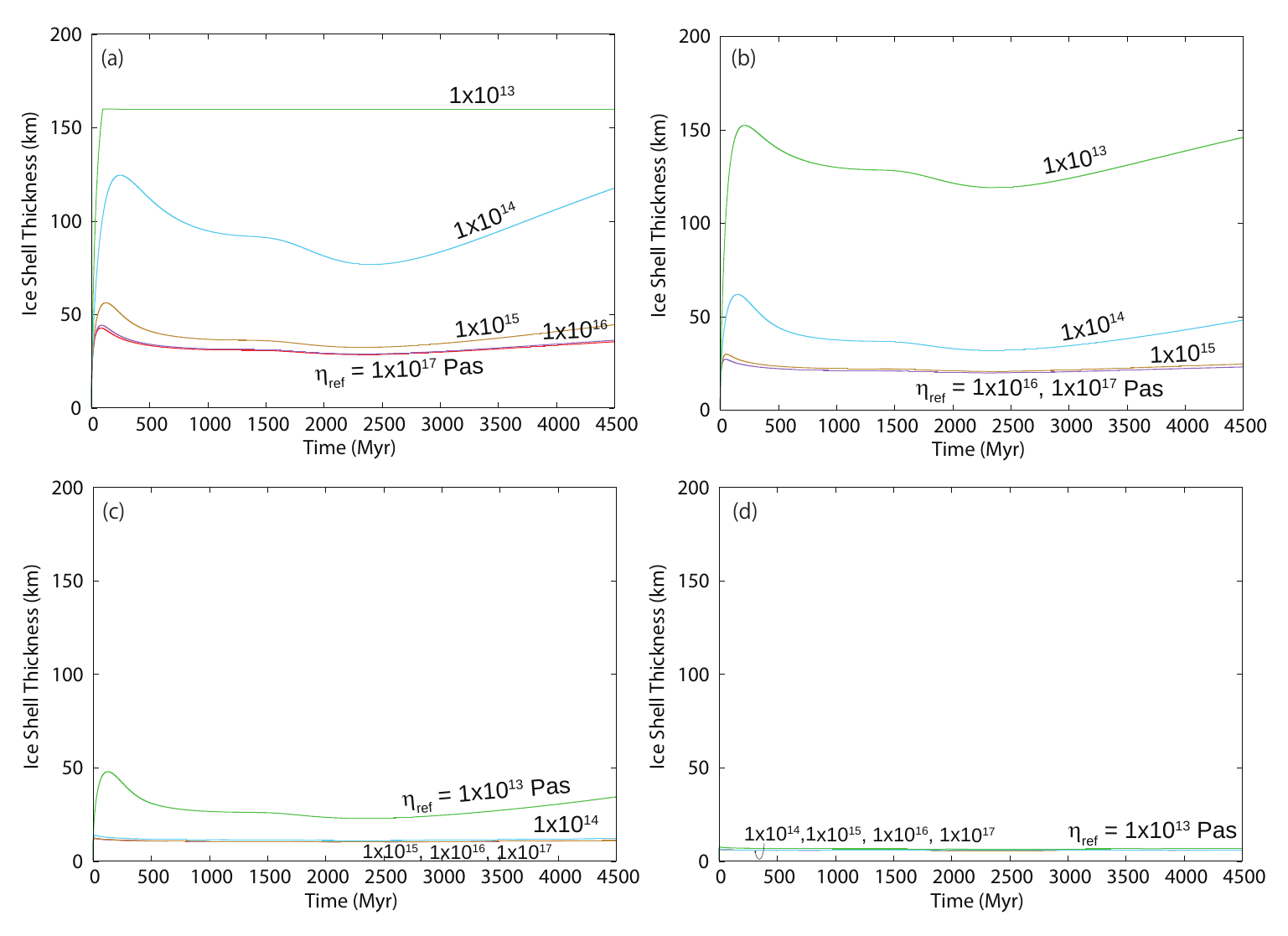}
\caption{Time evolution of the ice shell thickness for different ice reference viscosities ($\eta_{ref}$) under different parameter conditions.
D$_{H_{2}O}$\,=\,160\,km and $\rho_{core}$\,=\,6,500\,kg/m$^{3}$ (R$_{core}$\,=\,364\,km) are assumed.
Results for the tidal heating rate of (a) 10\,W/m$^{2}$, (b) 20\,mW/m$^{2}$, (c) 50\,mW/m$^{2}$ and (d) 100\,mW/m$^{2}$ are depicted (same setup as illustrated in Figs.\ref{fig:prof_s_CI} and \ref{fig:prof_s_CI_etaref}).}
\label{fig:shell_thickness_CI_tide}
\end{figure*}

Figure\,\ref{fig:shell_thickness_CI_tide} depicts the time evolution of the ice--shell thickness, including various tidal heating values between 0 and 100\,mW/m$^{2}$.
Under the assumptions of D$_{H_{2}O}$\,=\,160\,km and $\rho_{core}$\,=\,6,500\,kg/m$^{3}$ (R$_{core}$\,=\,364\,km), a model with an ice--shell thickness of 160\,km suggests that the moon does not possess a sub--surface ocean.
For the tidal heating rate of 10\,mW/m$^{2}$ (Fig.\,\ref{fig:shell_thickness_CI_tide}a), the current ice--shell thickness will be between 35 and 45\,km if the $\eta_{ref}$ is greater than 1$\times10^{15}$ Pa\,s, which is consistent with the previous estimates \citep{howell21}.
A larger tidal heating rate leads to a thinner ice shell.
For 20\,mW/m$^{2}$ (Fig.\,\ref{fig:shell_thickness_CI_tide}b), the subsurface ocean can be presently sustained for all values of the ice viscosities $\eta_{ref}$, although for a smaller ice viscosity of $\eta_{ref}=$10$^{13}$\,Pa\,s, the ice shell will be too thick compared with the previous estimates.
For 50\,mW/m$^{2}$ (Fig.\,\ref{fig:shell_thickness_CI_tide}c), the ice shell has a thickness of approximately 10\,km excepting the case of the ice reference viscosities of 1$\times10^{13}$ Pa\,s.
The ice shell has a thickness of 5\,km at a maximum heating rate of 100\,mW/m$^{2}$ for all ice reference viscosities (Fig.\,\ref{fig:shell_thickness_CI_tide}d), which implies that a 3--4\,km thickness that has been previously estimated based on the impact craters morphology \citep{turtle01} needs such a large tidal heating rate when the crater was formed.

\begin{figure}
\centering
\includegraphics[scale=0.15]{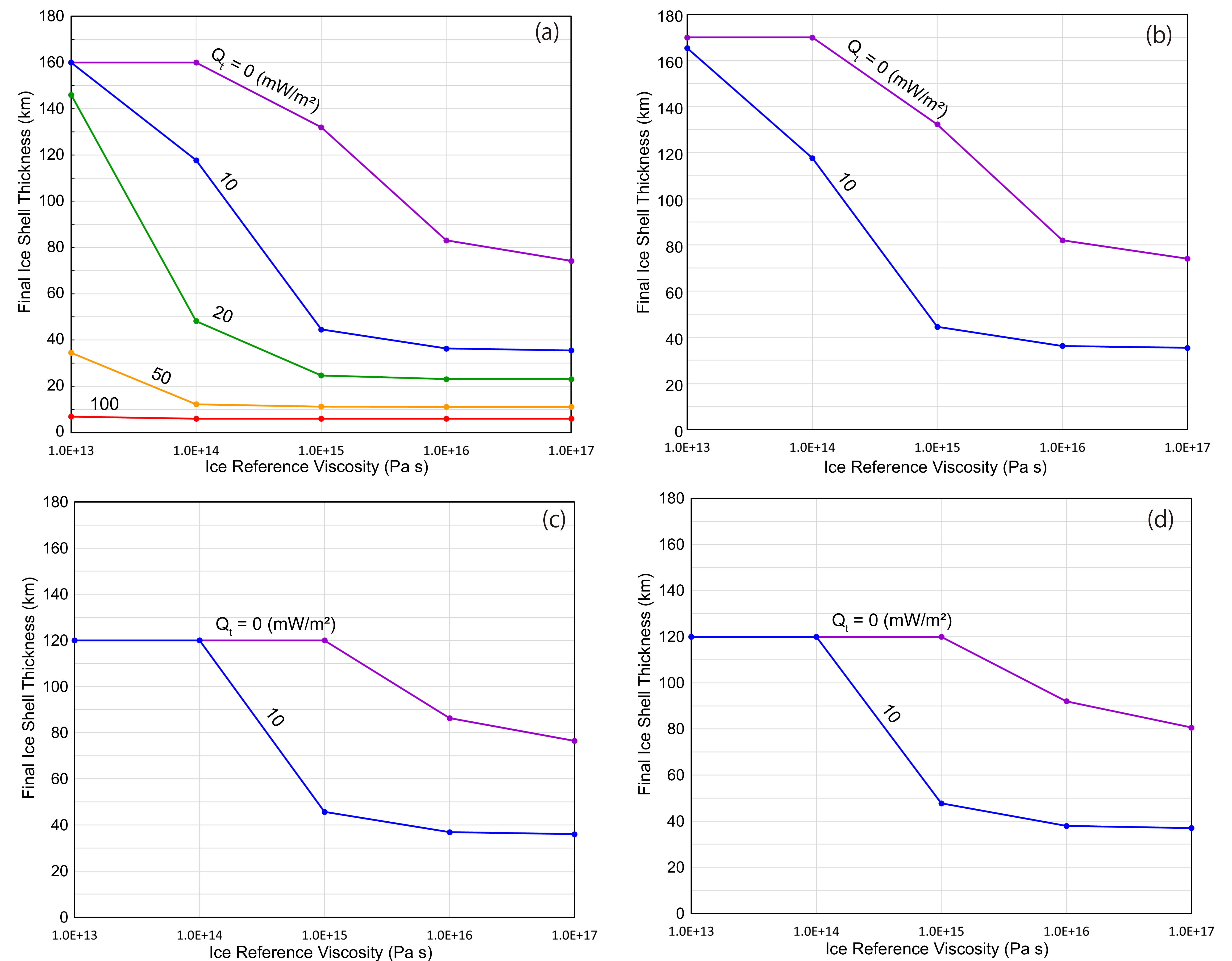}
\caption{Thickness of the ice shell at present as a function of the ice reference viscosity under different tidal heating rate.
Results for the case of 
(a) D$_{H_{2}O}$\,=\,160\,km and $\rho_{core}$\,=\,6,500\,kg/m$^{3}$ (R$_{core}$\,=\,364\,km, same structural settings as Fig. \ref{fig:shell_thickness_CI_tide}), 
(b) D$_{H_{2}O}$\,=\,170\,km and $\rho_{core}$\,=\,8,000\,kg/m$^{3}$ (R$_{core}$\,=\,178\,km), 
(c) D$_{H_{2}O}$\,=\,120\,km and $\rho_{core}$\,=\,8,000\,kg/m$^{3}$ (R$_{core}$\,=\,499\,km), 
(d) D$_{H_{2}O}$\,=\,120\,km and $\rho_{core}$\,=\,5,500\,kg/m$^{3}$ (R$_{core}$\,=\,678\,km),
are shown.}
\label{fig:final_thickness_CI_tide}
\end{figure}

Figure\,\ref{fig:final_thickness_CI_tide} depicts the final ice shell thickness as a function of the ice reference viscosity and the tidal heating rate for different interior structure values D$_{H_{2}O}$ and $\rho_{core}$.
Figure\,\ref{fig:final_thickness_CI_tide}a corresponds to the results presented in Fig.\,\ref{fig:shell_thickness_CI_tide}.
In the absence of tidal heating ($Q_{t}$\,=\,0\,mW/m$^{2}$), the 3D numerical study \citep{vilella20} suggests that the final shell thickness is $\le$\,90\,km if the ice reference viscosity is more than 1$\times10^{16}$ Pa\,s.
In case of the tidal heating of 10\,mW/m$^{2}$ and 20\,mW/m$^{2}$, the shell thickness is $\le$\,90\,km if the $\eta_{ref}$ is $\ge$\,1$\times$10$^{15}$ and 1$\times$10$^{14}$ Pa s, respectively.
Regardless of the $\eta_{ref}$, if the tidal heating is $\ge$\,50\,mW/m$^{2}$, the shell thickness will be $\le$\,40\,km.
These results are consistent with the other estimates of the ice--shell thickness based on the current steady-state heat balance \citep{howell21}, and the required rate of tidal heating is a reasonable value for the theoretical predictions \citep{hussmann02,tobie03,sotin09}.

In the case of the smallest core (R$_{core}$\,=\,178\,km) (Fig.\,\ref{fig:final_thickness_CI_tide}b), the overall trend in the results is mostly the same as that shown in Fig.\,\ref{fig:final_thickness_CI_tide}a, with the exception that the maximum shell thickness is slightly larger because the thickness of the hydrosphere (D$_{H_{2}O}$) is 10\,km larger than in the previous case.
For the smallest D$_{H_{2}O}$ of 120\,km, different core sizes do not significantly affect the final ice shell thickness (Fig.\,\ref{fig:final_thickness_CI_tide}c and \ref{fig:final_thickness_CI_tide}d).
In the absence of tidal heating, the difference in shell thickness for the various values of D$_{H_{2}O}$ and $\rho_{core}$ is less than a few kilometres.
If the tidal heating is $\ge$\,10\,mW/m$^{2}$, the shell thickness is generally the same regardless of the values of D$_{H_{2}O}$ and $\rho_{core}$ because the heat flux from the rocky mantle is approximately 10\,mW/m$^{2}$.
Figure \,\ref{fig:final_thickness_CI_tide} demonstrates that the ice--shell thickness is mainly controlled by the ice reference viscosity and the tidal heating rate.
It should be note that the above results are based on a simple calculation model assuming a constant tidal heating rate.
Additional calculations incorporating the coupled thermal-orbital evolution using a more realistic model are required for a further understanding \citep[e.g.,][]{hussmann04}.

\subsection{Effect of the radioactive isotope abundances}
\label{subsec:abundances}
\begin{figure}
\centering
\includegraphics[scale=0.65]{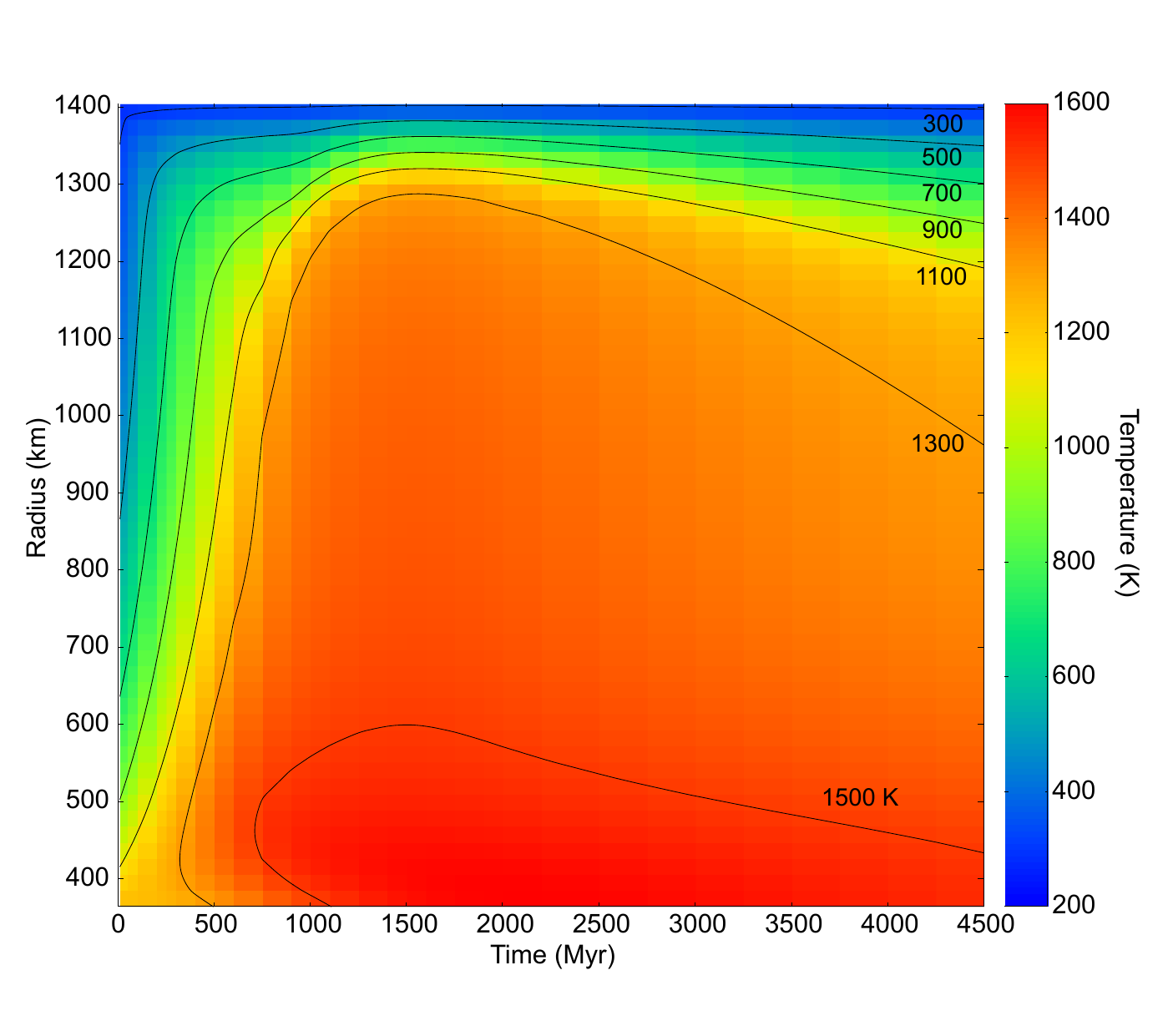}
\caption{Same as Fig. \ref{fig:prof_m_CI} but with ordinary chondritic abundances of long-lived radioactive isotopes.}
\label{fig:prof_m_OD}
\end{figure}

\begin{figure}
\centering
\includegraphics[scale=0.4]{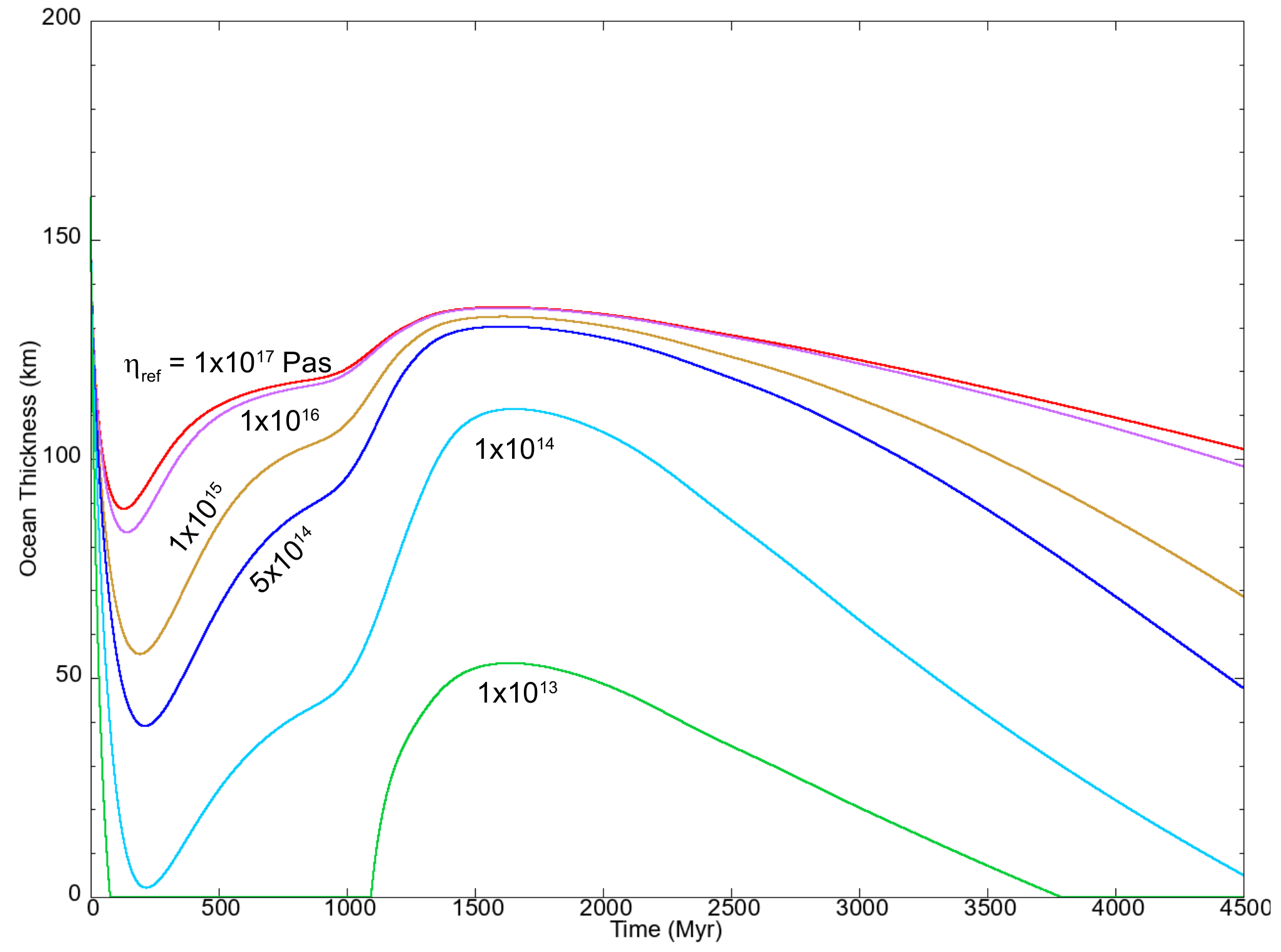}
\caption{Same as Fig. \ref{fig:prof_s_CI_etaref} but with the ordinary chondritic concentration in the rocky mantle.}
\label{fig:prof_s_OD_etaref}
\end{figure}

The ice--shell thickness is also determined by the concentration of the radioactive isotopes in the rocky mantle.
Figure\,\ref{fig:prof_m_OD} depicts the thermal evolution of the rocky mantle with the ordinary chondritic concentration of the long-lived radioactive isotopes.
In this case, the total amount of isotopes is approximately 30\% greater than the chondritic abundance; hence the temperature and heat flux increase in comparison to the previous results (Fig.\,\ref{fig:prof_m_CI}).
Figure\,\ref{fig:prof_s_OD_etaref} represents the evolution of ocean thickness over time using the same parameters as depicted in Fig.\ref{fig:prof_s_CI_etaref}, but with ordinary chondritic abundances in the rocky mantle.
The increased heat flux from the mantle results in a thicker ocean.
In the case of $\eta_{ref}=$1$\times$10$^{13}$\,Pa\,s, the ocean can be temporally formed; however, for $\eta_{ref}=$1$\times$10$^{14}$\,Pa\,s, the thin ocean can be sustained until the present day.

\begin{figure}
\centering
\includegraphics[scale=0.55]{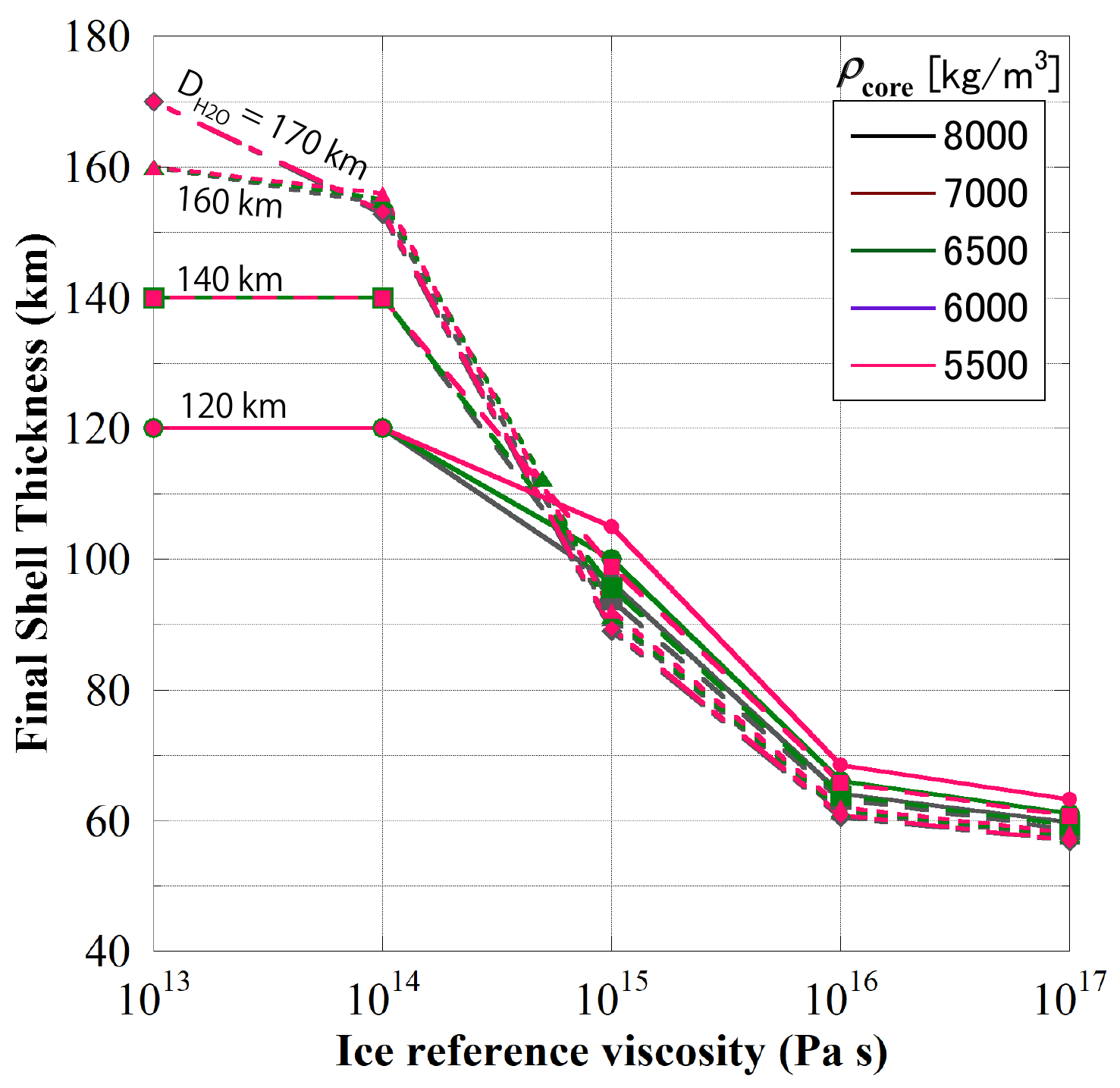}
\caption{Thickness of the ice shell at present as a function of the reference viscosity with ordinary chondritic concentrations under different structural conditions.}
\label{fig:final_thickness_OD}
Fig. \ref{fig:final_thickness_OD}.
\end{figure}

Figure\,\ref{fig:final_thickness_OD} depicts the final thickness of the ocean and the ice shell under computation settings with an ordinary chondritic concentration in the mantle.
A larger amount of radioactive heat source results in 15\,--\,20\,km thicker ocean than in the case of CI chondritic abundances (Fig.\,\ref{fig:final_thickness_CI}).
Thus, the final thickness of the ice shell is $\le$\,55\,km for all $\eta_{ref}$.
Nevertheless, $\eta_{ref}$ must be $\ge$\,1$\times$10$^{15}$\,Pa\,s for the ice shell thickness to be less than the previous estimates of 90\,km.

\begin{figure}
\centering
\includegraphics[scale=0.45]{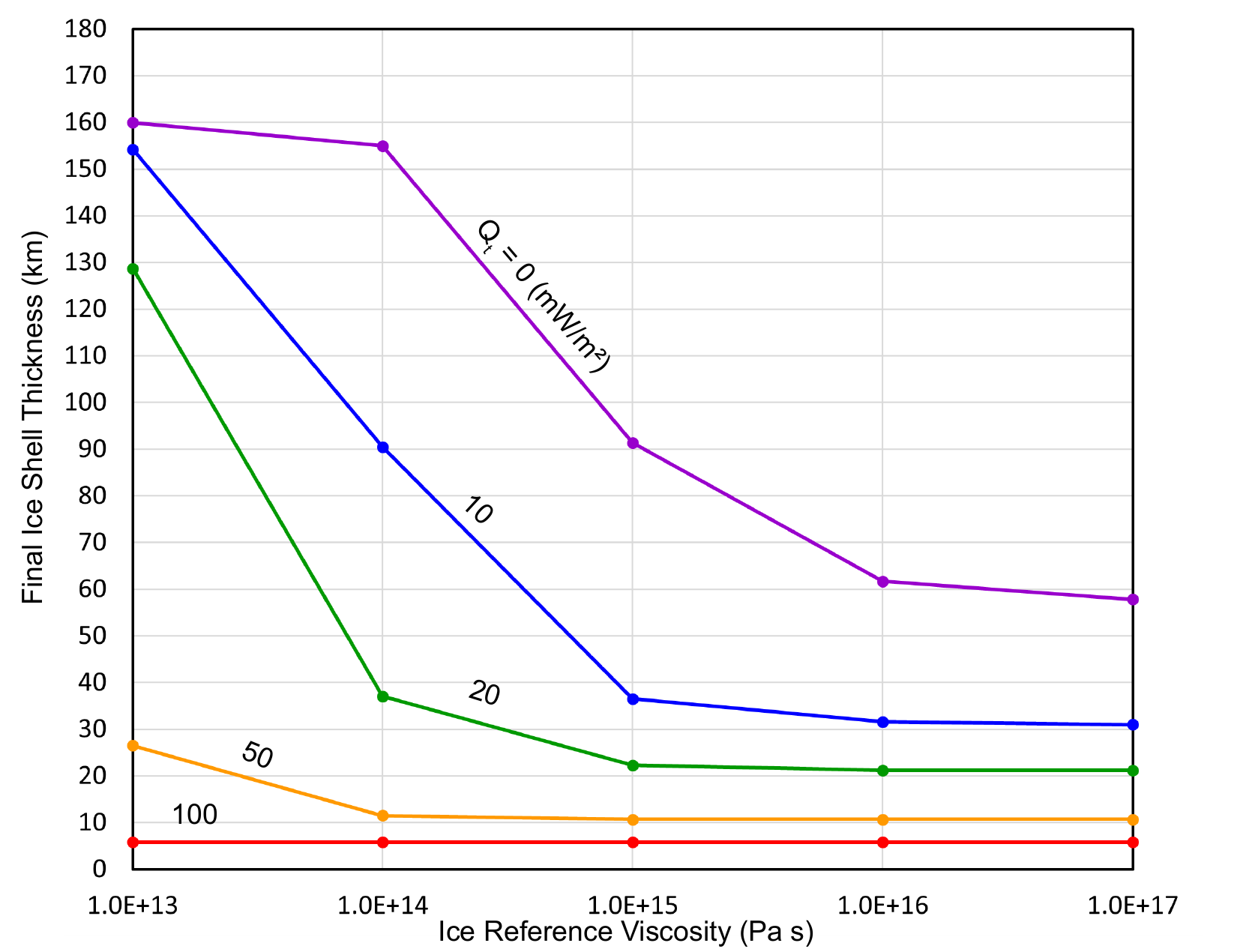}
\caption{Thickness of the ice shell at present as a function of the ice reference viscosity and the tidal heating for the case of D$_{H_{2}O}$\,=\,160\,km (same as Fig. \ref{fig:final_thickness_CI_tide}a) with ordinary chondritic concentration in the core.}
\label{shell_thickness_OD_with_tide}
\end{figure}

Figure\,\ref{shell_thickness_OD_with_tide} depicts the final thickness of the ice shell as a function of the ice reference viscosity and the tidal heating rate for the case of D$_{H_{2}O}$\,=\,160\,km and $\rho_{core}$\,=\,6,500\,kg/m$^{3}$ (R$_{core}$\,=\,364\,km), with the ordinary chondritic concentration in the rocky mantle.
In the case of the tidal heating of 10\,mW/m$^{2}$ and 20\,mW/m$^{2}$, the shell thickness is $\le$\,90\,km if the $\eta_{ref}$ is more than 1$\times$10$^{14}$ and 2$\times$10$^{13}$\,Pa\,s, respectively.
\subsection{Capability of the core dynamo activity and conceivable constraints of the internal structure}
\begin{figure}
\centering
\includegraphics[scale=0.45]{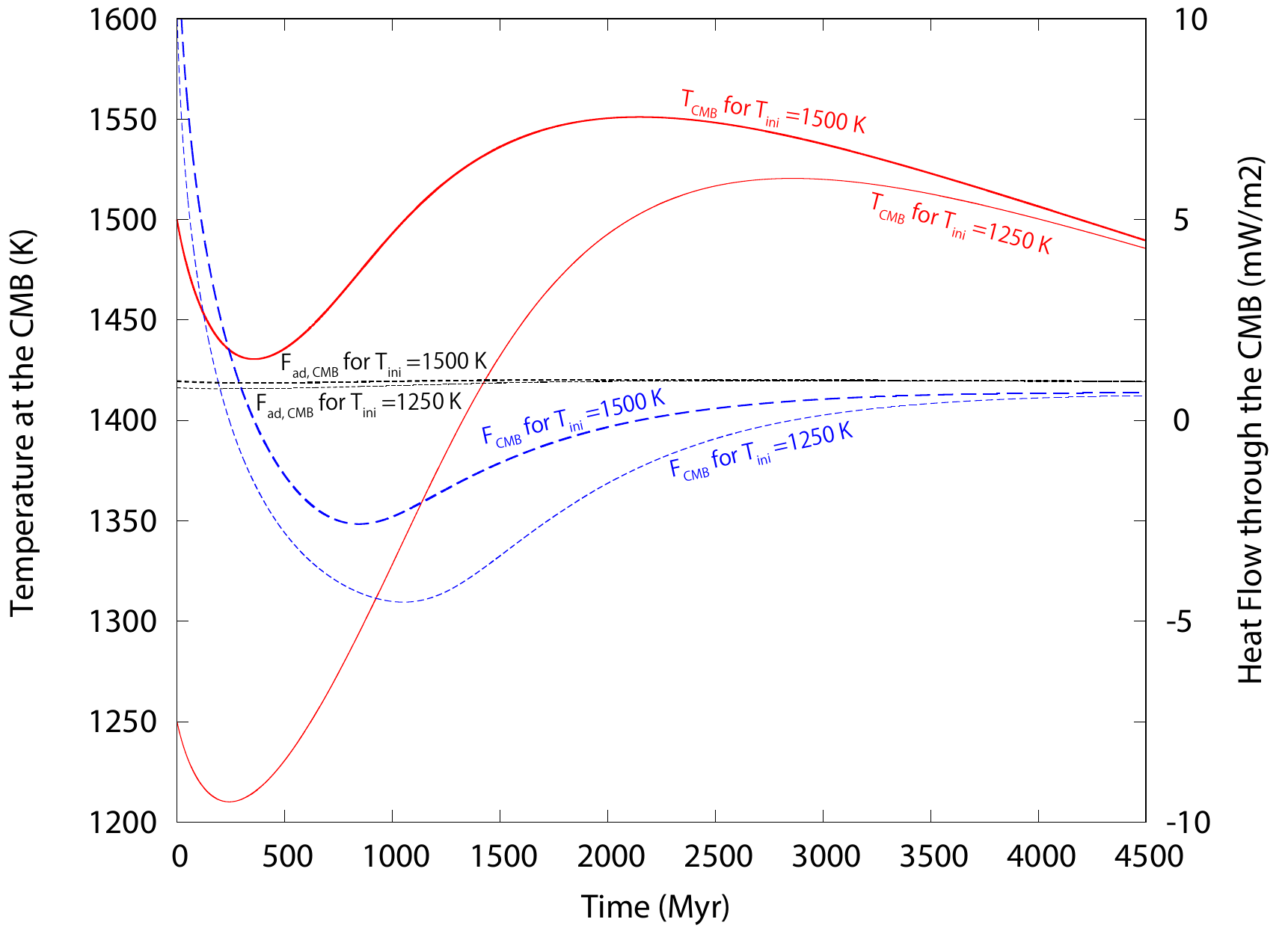}
\caption{Temporal changes of the temperature at the CMB (solid red curves) and heat flux across the CMB (dashed blue curves).
Thin and thick lines indicate the case for the initial temperature at the CMB of 1,250\,K and 1,500\,K, respectively.
Dotted black lines indicate the adiabatic heat flux defined by the temperature at the CMB ($F_{CMB}$\,=\,$k_{c}\alpha_{c} gT_{CMB}/C_{p,c}$).
Results for the case of D$_{H_{2}O}$\,=\,160\,km and $\rho_{core}$\,=\,6,500\,kg/m$^{3}$ (R$_{core}$\,=\,364\,km) with the CI chondritic abundance of the radioisotopes in the rocky mantle, which is same calculation setting as Figs. \ref{fig:prof_s_CI}, \ref{fig:prof_m_CI} and \ref{fig:prof_s_CI_etaref}.}
\label{fig:Tcmb_Fcmb}
\end{figure}

Figure\,\ref{fig:Tcmb_Fcmb} depicts the temporal change in the CMB temperature and heat flux over 4.5\,Gyr, assuming D$_{H_{2}O}$\,=\,160\,km and $\rho_{core}$\,=\,6,500\,kg/m$^{3}$ (R$_{core}$\,=\,364\,km).
During the early stages, after approximately 0.5\,Gyr, the temperature begins to increase due to the decay of long-lived radioactive isotopes in the mantle.
The metallic core is heated by the overlying mantle and it remains in the stable stratification state in this stage.
When radioactive isotopes eventually decay, which takes a few billion years, the temperature in the deepest region of the mantle starts decreasing (see also Fig.\,\ref{fig:prof_m_CI}).
The heat flux across the CMB increases as the rocky mantle cools down.
The core density used in this calculation corresponds to approximately 22\,wt\% of the sulfur content in the Fe--FeS system, and the solidus temperature is approximately 1,390\,K.
The metallic core may currently be at least partially molten as the temperature at the CMB may be higher than the solidus at both latter stages and the present for both initial conditions.
In contrast, the heat flux across the CMB cannot exceed the adiabatic heat flux, and thus the thermal convection cannot be driven even if the core is molten.
Therefore, I considered that this structure satisfies the melting condition but not the cooling condition, and the current core is incapable of generating an active dynamo, which is consistent with the absence of the intrinsic magnetic field originating from the core.
As a result, this interior structure can be considered as a candidate of the current conditions of Europa.

\begin{figure}
\centering
\includegraphics[scale=0.3]{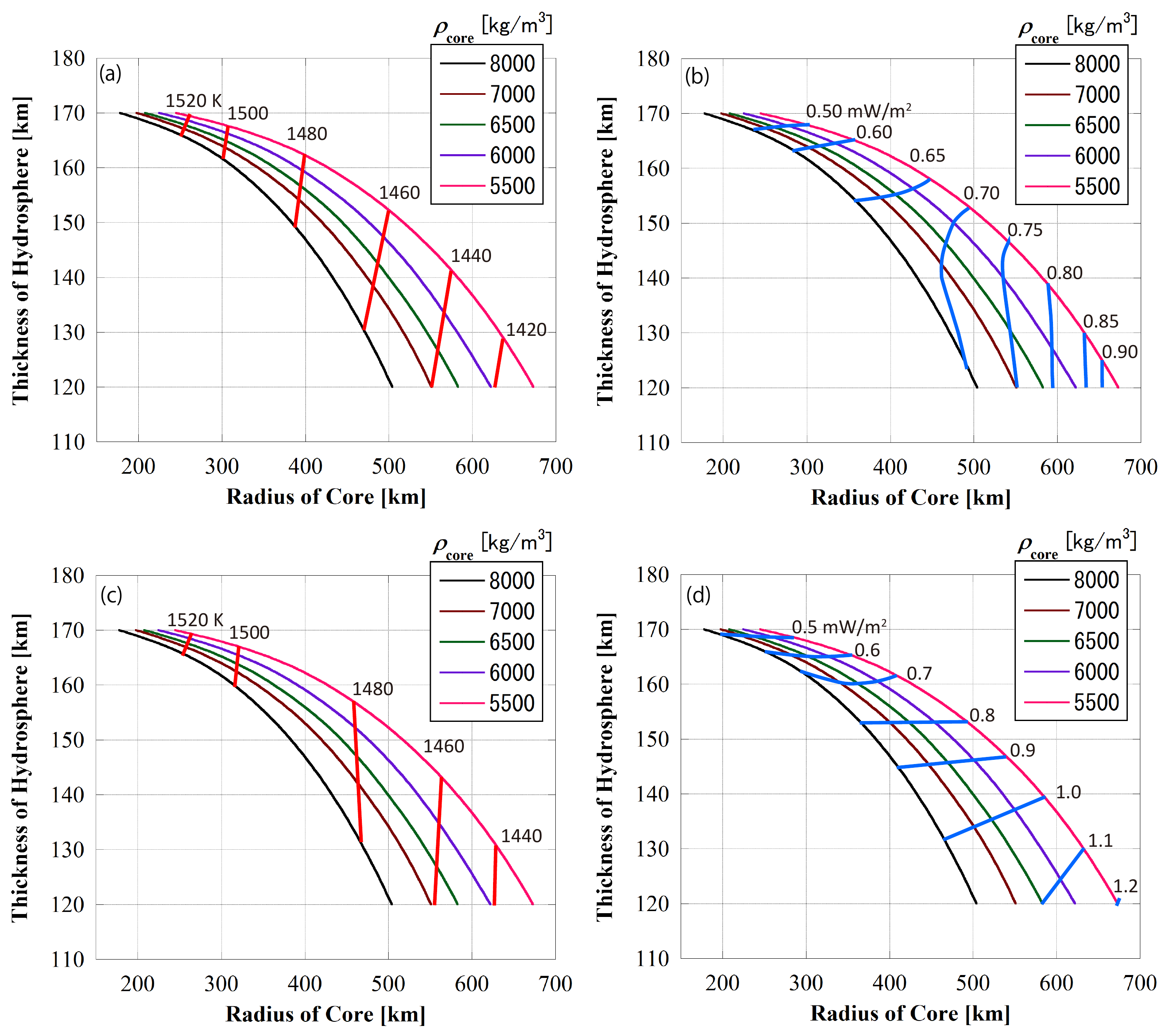}
\caption{Contours of the temperature at the CMB (a and c) and the heat flux across the CMB (b and d) after 4.5 Gyr from the beginning of the calculations.
Upper figures (a and b) and lower figures (c and d) represent the case for the initial temperature at the CMB of 1,250\,K and 1,500\,K, respectively.}
\label{fig:tcmb_fcmb_contour}
\end{figure}

Comprehensive calculations were performed on various interior structures to explore the optimal condition for a dynamo at the CMB within 4.5\,Gyr.
Figure\,\ref{fig:tcmb_fcmb_contour} represents the temperature at the CMB (a and c) and the heat flux across the CMB (b and d) after 4.5\,Gyr for initial temperatures at the CMB of 1,250\,K and 1,500 \,K.
A smaller core results in a higher temperature at the CMB because of a larger amount of the radioactive heat sources in the mantle.
In contrast, a larger core results in a lower temperature at the CMB and a greater heat flux across the CMB.
In the case of a larger core, the heat flux across the CMB depends on the size of the core (heat capacity).
In the case of a smaller core, a thinner hydrosphere (thicker rocky mantle and a large amount of radiogenic heat source) leads to greater heat flux across the CMB.
For higher initial temperatures, the thickness of the hydrosphere determines the cooling efficiency of the core.
If the temperature and heat flux for a certain interior structure, i.e. specific combination of $D_{H_{2}O}$, $R_{core}$ and $\rho_{core}$, can exceed the solidus and the adiabatic flux, then such a structure can be considered to satisfy the melting and cooling conditions, respectively.
If both conditions are satisfied, such a structure is capable of driving the dynamo activity (thermal convection) in the metallic core. 

\begin{figure}
\centering
\includegraphics[scale=0.3]{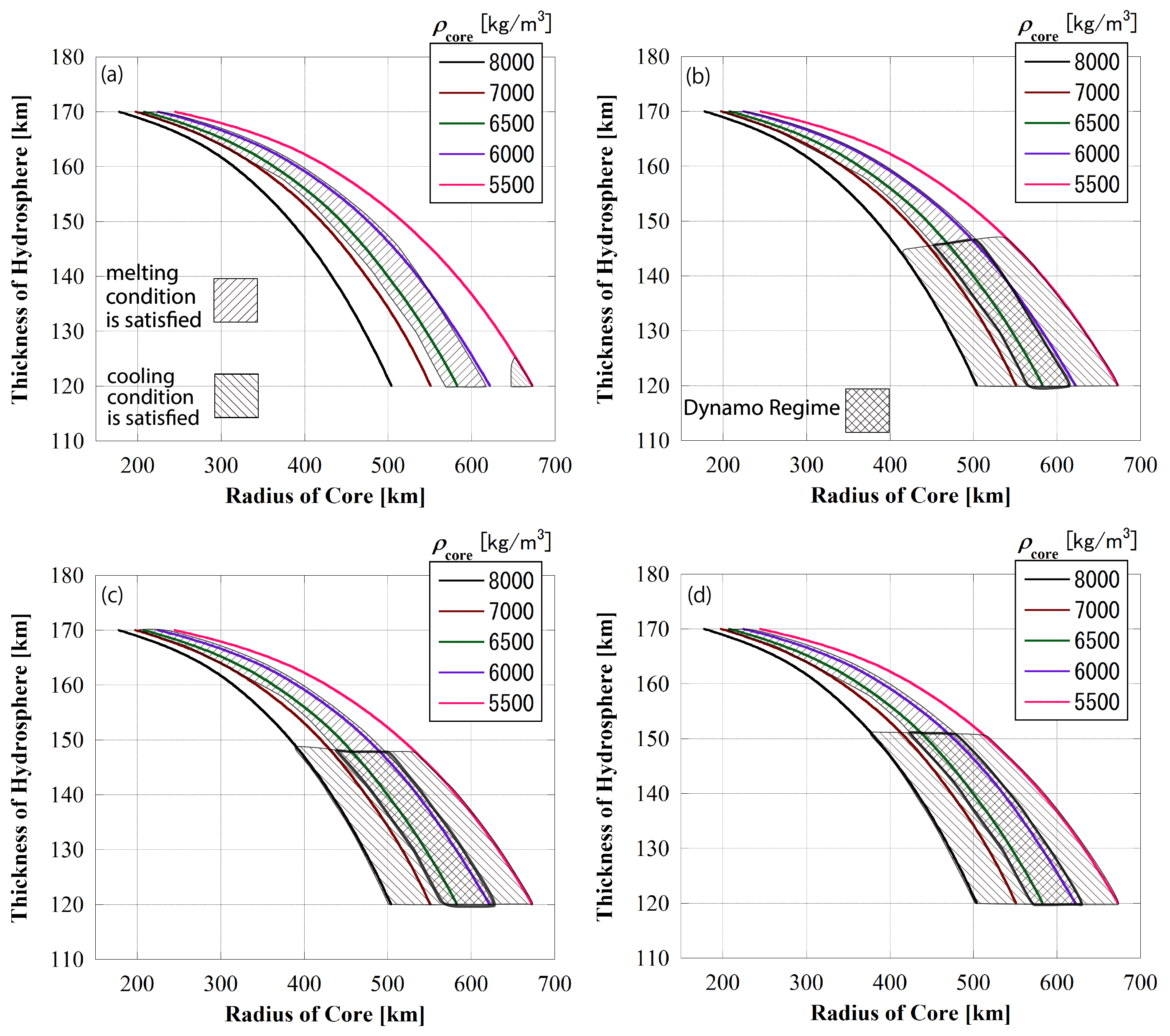}
\caption{Admissible range of structural conditions for a driving core dynamo activity.
(a) and (b) assume the CI chondritic abundances of the radioactive isotopes in the mantle,
and (c) and (d) assume the OD chondritic abundances.
(a) and (c) assume the initial temperature at the CMB of 1,250\,K, while (b) and (d) assume 1,500\,K.
Each hatched area indicates the structural region where the melting and cooling conditions are satisfied, respectively.
The melting condition is that the temperature at the CMB is higher than the melting point of the assumed core composition, implying that the metallic core is at least partially molten.
The cooling condition is that the heat flux at the CMB must also be greater than the adiabatic temperature gradient.
Overlapping both hatched areas indicates a specific range of the structures that are capable of driving the dynamo denoted as the Dynamo Regime.
}
\label{fig:dynamo_regime}
\end{figure}

The diagram of structural settings relating to the condition for dynamo activity in the core is depicted in Figure\,\ref{fig:dynamo_regime}.
It illustrates the combination of structural parameters that can satisfy the heating condition (current CMB temperature can exceed the solidus for the assumed core density) and the cooling condition (current heat flux across the CMB can exceed the adiabatic flux).
The specific region which satisfies both conditions is identified as the {\it dynamo Regime}.
The solidus temperature is higher when the bulk core composition is close to the end members of the Fe--FeS system, making it difficult to satisfy the melting condition in all cases.
Therefore, the dynamo regime also varies with variations in the core's sulfur content.
For an initial T$_{CMB}$\,=\,1,250\,K with CI chondritic abundances (Fig.\,\ref{fig:dynamo_regime}a), there is no dynamo regime region within the range of the interior structure. 
The structural range where the cooling condition can be satisfied increases for a higher initial temperature of T$_{CMB}$\,=\,1,500\,K, and the dynamo regime can be observed (Fig.\,\ref{fig:dynamo_regime}b).
According to the expected sulfur amount in the core from the bulk CI chondritic composition, a higher density (smaller sulfur amount) core is unlikely \citep{bercovici22}.
In the case of ordinary chondritic abundances (Fig.\,\ref{fig:dynamo_regime}c and d), the dynamo regime can also emerge regardless of the initial temperature.
The metallic core should be smaller, closer to the endmembers of the Fe--FeS system in composition, have a lower initial temperature, have a thicker hydrosphere and have lower radioactive isotope abundances in order for there to be no trace of dynamo activity in the current Europa.
It should be noted that a lower density (larger sulfur amount) core is unlikely according to the expected sulfur amount in the core from the bulk ordinary chondritic composition  \citep{bercovici22}.

\section{Discussions}
\label{sec:discussions}
%
\subsection{Effect of compositional convection}
In this study, only thermal convection is considered for the dynamics of Europa's metallic core of Europa.
In the terrestrial case, cooling and solidification of the liquid core results in the growth of the inner core, followed by the induction of convection in the outer core.
This process serves as an energy source for the generation of the intrinsic magnetic field, as well as the thermal convection.
Depending on the thermal state, Europa's core may segregate into an outer and an inner core, and bulk sulfur content will determine the composition and structure within the core.
The eutectic temperature of the Fe--FeS system decreases with increasing pressure up to 14 GPa \citep{fei97}, while the pressure range in Europa's core is between 3 and 4\,GPa and 6\,GPa, depending on the core density and size.
The sulfur content at the eutectic decreases with increasing pressure.

Denser iron (Fe) will sink if the sulfur content is lower than the eutectic composition.
Fe snow forms at the CMB as the core cools, after which Fe sinks and the sulfur content increases. 
A higher sulfur content reduces the liquidus, which causes Fe to remelt in a deeper region and cause chemical convection. 
A solid Fe inner core will eventually form as sinking Fe approaches the centre \citep{breuer15}.
According to \citet{hauck06}; and \citet{zhan12}, the dissipated power by Fe snow is sufficient to drive a dynamo.
However, \citet{bland08} pointed out that compositional convection is constrained by the latent heat that is quickly released from the area surrounding the CMB.
\citet{ruckriemen18} suggested that the remelting of Fe below the snow zone and the release of buoyancy produce convection which may drive the dynamo.
All of these studies assume a hot start, mantle and core initial temperatures of approximately 2,000\,K, and search conditions to maintain an initially activated dynamo until the present day.

If the higher sulfur content exceeds the eutectic composition, the solid outer core will be composed of the lighter FeS.
However, such sulfur--rich side of the eutectic has received less attention.
It remains to be seen whether sufficient power can be generated for a dynamo in this scenario, which is beyond the scope of this study.

At least melting conditions need to be satisfied even for compositional convection (core crystallisation).
However, even if the cooling heat flux does not exceed the adiabatic heat flux, the crystallisation and associated compositional convection may occur depending on the core composition.
In this respect, the cooling conditions in this study are a stringent criterion, and the core solidification could extend the dynamo regime (i.e. the cooling condition could be easily satisfied).
The cooling rate controls the crystallisation rate.
This study does not consider the core crystallisation because of the lack of observational information and constraints for the presence of structure in the core.
However, if such information becomes available in the future, the cooling conditions could be constrained more strictly.
\subsection{Effect of the rock rheology}
\label{subsec:rock_rheology}
\begin{figure}
\centering
\includegraphics[scale=0.35]{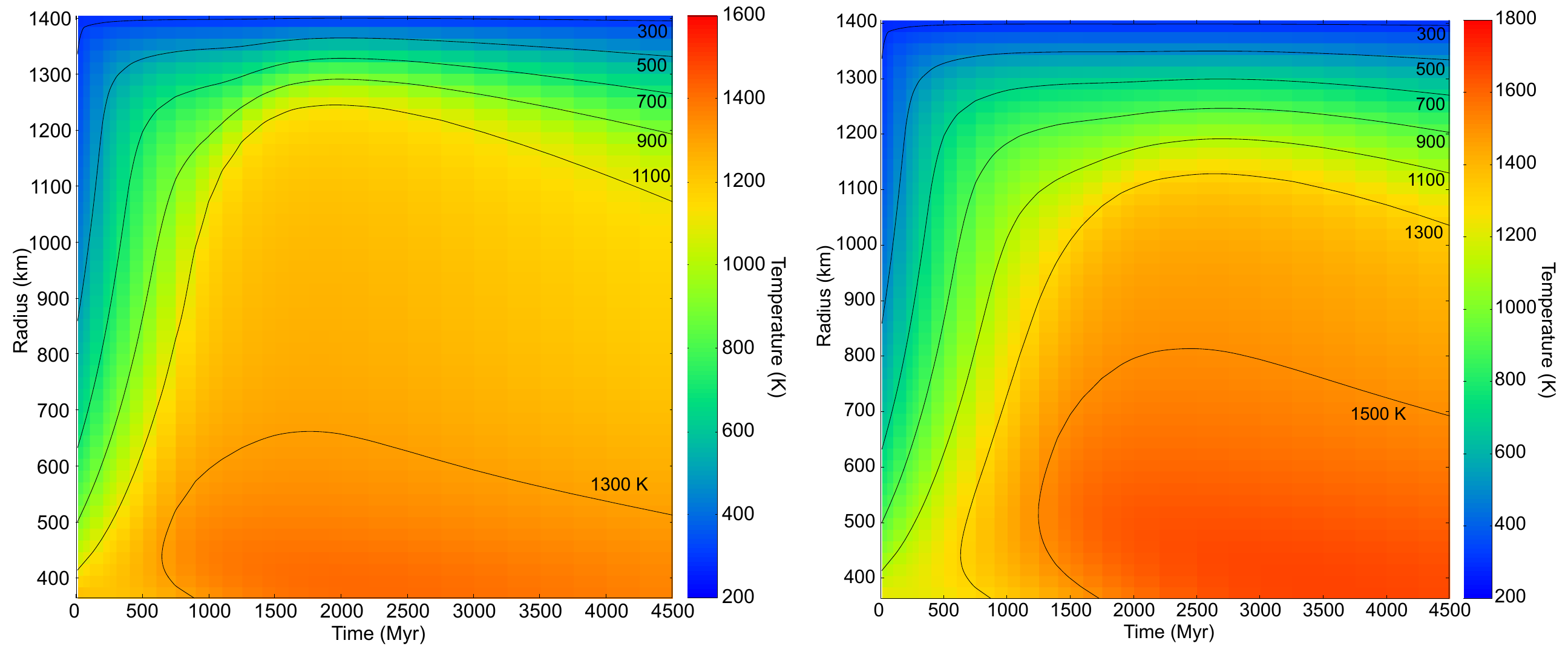}
\caption{Same as Fig.\ref{fig:prof_s_CI_etaref} but with different values of rock viscosity, $A=21.0$ (left), and $A=26.5$ (right).}
\label{fig:prof_m_A}
\end{figure}

For the diffusion creep rheology for the rocky mantle which is assumed in this work, various values of $A$ in the equation (\ref{eq:rock_viscosity}) in range between 21.0 and 26.5 have been suggested in the previous experimental works \citep{karato86,hirth03}.
As a references, for $T=1,000\,K$, $\eta_{rock}$\,=\,1.9$\times$10$^{23}$\,Pa\,s for $A=21.0$, $\eta_{rock}$\,=\,7.0$\times$10$^{24}$\,Pa\,s for $A=23.25$ (regarded as a typical value here) and $\eta_{rock}$\,=\,1.3$\times$10$^{27}$\,Pa\,s for $A=26.5$.
Figure \ref{fig:prof_m_A} shows the thermal evolution of the rocky mantle in the case of $A$\,=\,21.0 and $26.5$.
Because of the different dependence of viscosity on the temperature, the deeper region of the rocky mantle is about 200 K lower and higher for $A$\,=\,21.0 and 26.5, respectively, compared with the typical value of $A$\,=\,23.25 shown in Figure\,4.
Alternatively, near the upper boundary of the mantle, the temperature profile is almost similar and the difference of the heat flux at the surface of the mantle among A values is small, therefore the difference of the resulting ice-shell thickness at present is only $\le$\,2\,km.
\subsection{Effect of a higher value for the moment of inertia factor of Europa}
\label{subsec:moi_03547}
\begin{figure}
\centering
\includegraphics[scale=0.3]{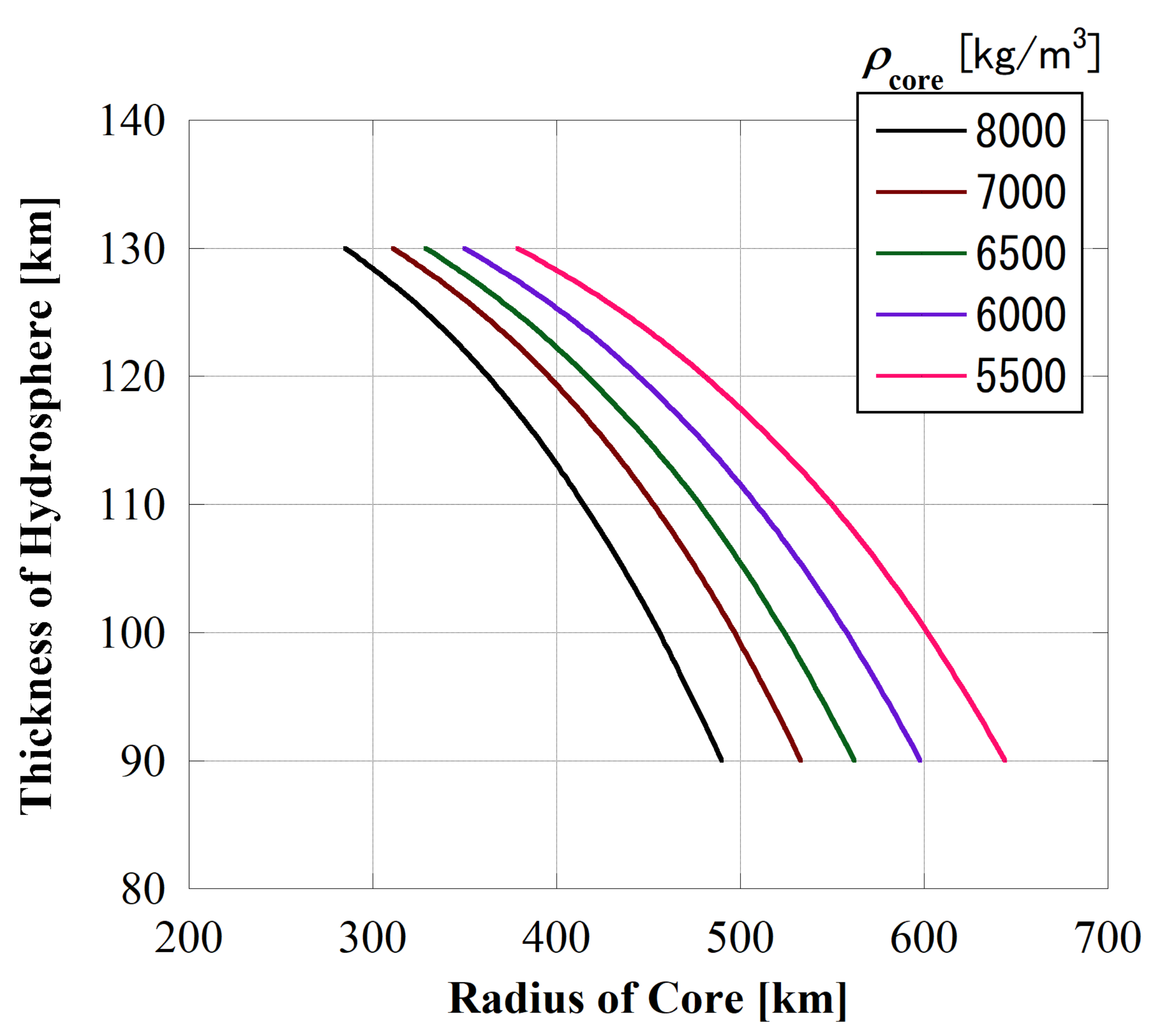}
\caption{Same as Figure \ref{fig:interior} but based on the moment of inertial factor of 0.3547 \citep{casajus21}.
}
\label{fig:interior_moi03547}
\end{figure}

A recent re-investigation of the Galileo gravity data yielded a higher value for the moment of inertia factor, 0.3547\,$\pm$\,0.0024, suggesting that Europa may have a thinner hydrosphere \citep{casajus21}.
Figure\,\ref{fig:interior_moi03547} represents the possible range of constant-density shell models for the Europa's interior.
The thickness of the hydrosphere is approximately 30--40\,km thinner, and the density of the rocky mantle is smaller  than previous estimates as \citet{casajus21} suggested, if Europa has a 3-layered (hydrosphere, rocky mantle, and metallic core) structure.
The size of the metallic core is almost similar to former estimates shown in Figure\,\ref{fig:interior}  based on the previous value of the moment of inertia of 0.346. 
The thickness of the mantle is larger according to the thinner hydrosphere, so that the difference in the overall mass of the mantle is only a few weight per cent. 
Therefore, the thermal evolution of the mantle and the core is not likely to change much. 
Alternatively, since the hydrosphere is thinner than previously estimated, the lifetime of the ocean is expected to be about 1\,Gyr shorter, even if the heat from the mantle and the melting point viscosity of the ice are similar. 
Detailed quantitative evaluation is useful for future work.
\subsection{Perspective for future spacecraft missions}
In 2030, Europa Clipper will flyby to Europa carrying multiple scientific instruments to reveal the surface, interior and surrounding environment \citep[e.g.][]{howell20}. 
Because of multiple flybys including high latitude passes, the Europa Clipper will determine J$_{2}$ and C$_{22}$ independently (without an assumption for hydrostatic equilibrium).
Detectability of 1\% departure from hydrostatic behavior would bring 10 times more accurate than the recent estimate by \citet{casajus21}, and would narrow the space of possible metallic core radius and bulk rock density \citep{mazarico23}.
The gravity signal at long wavelengths is likely to reflect effects from the ocean-rock interface, and gravity anomalies arising from the sea-floor will be compensated at longer wavelengths by the rocky mantle.
Assuming that this compensation and thus the magnitude of gravity anomalies that the Europa Clipper may detect can be related to the heat flow from the rocky mantle, we might distinguish between the radiogenic heat and the tidal heating depending on rock strength \citep{dombard19}.
Although a measurement of the tidal potential Love number k$_{2}$ with a small uncertainty (0.014\,--\,0.018) can provide one constraint on Europa's shell thickness and rigidity, the tradeoff between the shell thickness and rigidity will remain.
Thus, even if the Europa Clipper measures the displacement Love number h$_{2}$, its uncertainty will be greater than k$_{2}$, making it difficult to determine the ice--shell thickness \citep{mazarico23}.

Europa Clipper Magnetometer (ECM) determines the direction and strength of an induced field for at least two frequencies with a precision of $\pm$\,1.5\,nT, and reinvestigates the presence or the absence of the core--driven field.
ECM measures variations at the orbital (85.2\,h, $\sim$20\,nT) and second harmonic of the synodic period (5.6\,h, $\sim$20\,nT) are expected to reveal a unique combination of ice shell thickness, ocean thickness, and ocean conductivity that fits the data \citep{roberts23}.
Additionally, the conductivity of the ocean will be inferred from ECM and PIMS observations \citep{vance23}.
In an idealized case, a combination of measurements of the static gravity (moment of inertia), tidal Love number, magnetic induction field and radar penetration provides full characterisation of the ice shell, which allow the mean ice--shell thickness, the thickness of the sub--surface ocean and the conductivity of the ocean to be determined independently with error bars of $\pm$\,30km for the ocean, $\pm$\,5km for the ice shell \citep{roberts23}.
REASON sounding measurements would constrain the ice--shell thickness although thick ($>$ several tens of km) shell could not be directly sensed.
Combining digital terrain model, estimation of the elastic thickness the ice shell would be allowed.
Since the elastic strength depends on the temperature structure of the ice shell, additional assumptions about the thermal state or direct measurement of the surface thermal state by E-THEMIS (Europa THermal EMission Imaging System) lead to further constraints on the ice-shell thickness \citep{roberts23}.

Furthermore, a lander mission could take place in the future \citep{hand17}.
Seismic investigations can determine the depth of the ocean from the time delay between ocean multiples in the primary wave coda of teleseismic Europa quakes with a magnitude of 3 that can be expected to occur between once per week and once per month \citep{stahler18}.
\section{Summary}
\label{sec:summary}
A numerical simulation of Europa's interior thermal history was performed to investigate the temporal change of the ocean thickness and to determine the permissible range where the core dynamo cannot be generated at present.
The ice reference viscosity, tidal heating rate and radioactive isotope abundance in the rocky mantle all play major roles in determining the thickness of the ice shell.
The thickness of the hydrosphere and the metallic core density have no bearing on the thickness of the ocean.
If the ice shell has a large viscosity, and particularly, if the ice reference viscosity, $\eta_{ref}$ is $\ge$\,5$\times$10$^{14}$ Pa\,s, the subsurface ocean can persist even in the absence of tidal heating.
However such resulting ocean thickness is considerably thin, and the ice shell thickness is significantly greater than that estimated by previous geologic analyses.
Assuming tidal heating of 10\,mW/m$^{2}$ and 20\,mW/m$^{2}$, the ice shell thickness is $\le$\,90\, km if the $\eta_{ref}$ is $\ge$\,1$\times$10$^{15}$ and 1$\times$10$^{14}$ Pa\,s, respectively.
Regardless of the value of $\eta_{ref}$, if the tidal heating is $\ge$\,50\,mW/m$^{2}$, the shell thickness is $\le$\,40\,km.
This thickness is consistent with the previous estimates by geological and numerical approaches.
For ordinary chondritic abundances of the radioactive isotopes in the rocky mantle, the larger amount of heat source in the rocky mantle leads to a thicker ocean (a thinner ice shell) compared with the CI chondritic abundances.
In particular, for the tidal heating of 10\,mW/m$^{2}$ and 20\,mW/m$^{2}$, the ice shell thickness is $\le$\,90\,km if the $\eta_{ref}$ is $\ge$\,1$\times$10$^{14}$ and 2$\times$10$^{13}$\,Pa\,s, respectively.

Additionally, further analyses were conducted on the permissible range of the interior structure (hydrosphere thickness, metallic core density and radius) that lead to the absence of the core dynamo.
As a result, the inferred range of interior structure was narrowed down, compared with that range only based on the moment of inertia.
For the absence of a core dynamo, a lower initial temperature, smaller amount of heat source and a core composition closer to the endmember for the Fe--FeS system is preferred.
For example, if the rocky mantle has a CI chondritic abundances of radioactive isotopes, any core composition and hydrosphere thickness allows the absence of the core dynamo if the initial temperature at the CMB is lower than 1,250\,K.
For the case of an initial CMB temperature of 1,500\,K, a core density higher than 7,000\,kg/m$^{3}$ (S\,$<$\,15\,wt\%) or lower than 6,000\,kg/m$^{3}$ (S\,$>$\,29\,wt\%) is preferred, or the hydrosphere thicker than 145\,km for the core composition closer to the eutectic one is suitable.
According to the expected sulfur amount in the core from the bulk CI chondritic composition, a lower density (larger sulfur amount) core is preferred.
If the rocky mantle has the ordinary chondritic radioactive abundances, a core density higher than 7,000\,kg/m$^{3}$ (S\,$<$\,15\,wt\%) or lower than 5,800\,kg/m$^{3}$ (S\,$>$\,32\,wt\%) is preferred, or the hydrosphere thicker than 150\,km for the core composition closer to the eutectic one is required for the absence of the core dynamo regardless of the initial CMB temperature.
According to the expected sulfur amount in the core from the bulk ordinary chondritic composition, a higher density (smaller sulfur amount) core is preferred.
The lower pressure of Europa's rocky mantle due to its thinner hydrosphere compared with that of Ganymede may facilitate heat transfer in the mantle, lowering its temperature and making dynamo motion more challenging.
\clearpage
\section*{Acknowledgments}
We thank anonymous reviewers for constructive comments which greatly improved the manuscript.
This study was supported by KAKENHI from the Japan Society for Promotion of Science (Grant No. JP22K03700, JP17K05635 and JP22740285).
\clearpage


\bibliographystyle{elsarticle-harv} 
\bibliography{jkbibfile}

\begin{thebibliography}{77}
\expandafter\ifx\csname natexlab\endcsname\relax\def\natexlab#1{#1}\fi
\providecommand{\url}[1]{\texttt{#1}}
\providecommand{\href}[2]{#2}
\providecommand{\path}[1]{#1}
\providecommand{\DOIprefix}{doi:}
\providecommand{\ArXivprefix}{arXiv:}
\providecommand{\URLprefix}{URL: }
\providecommand{\Pubmedprefix}{pmid:}
\providecommand{\doi}[1]{\href{http://dx.doi.org/#1}{\path{#1}}}
\providecommand{\Pubmed}[1]{\href{pmid:#1}{\path{#1}}}
\providecommand{\bibinfo}[2]{#2}
\ifx\xfnm\relax \def\xfnm[#1]{\unskip,\space#1}\fi
\bibitem[{Abe(1993)}]{abe93}
\bibinfo{author}{Abe, Y.}, \bibinfo{year}{1993}.
\newblock \bibinfo{title}{Thermal evolution and chemical differentiation of the
  terrestrial magma ocean}.
\newblock \bibinfo{journal}{In: Takahashi, E., Jeanloz, R., Rubie, D. (Eds.),
  Evolution of the Earth and PlanetsNature,American Geophysical Union,
  Washington DC.} , \bibinfo{pages}{41--54}.
\bibitem[{Anderson et~al.(1998)Anderson, Schubert, Jacobson, Lau, Moore and
  Sjogren}]{anderson98}
\bibinfo{author}{Anderson, J.D.}, \bibinfo{author}{Schubert, G.},
  \bibinfo{author}{Jacobson, R.A.}, \bibinfo{author}{Lau, E.L.},
  \bibinfo{author}{Moore, W.B.}, \bibinfo{author}{Sjogren, W.L.},
  \bibinfo{year}{1998}.
\newblock \bibinfo{title}{Europa's differentiated internal structure:
  Inferences from four galileo encounters}.
\newblock \bibinfo{journal}{Science} \bibinfo{volume}{281},
  \bibinfo{pages}{2019--2022}.
\bibitem[{Andersson and Inaba(2005)}]{andersson05}
\bibinfo{author}{Andersson, O.}, \bibinfo{author}{Inaba, A.},
  \bibinfo{year}{2005}.
\newblock \bibinfo{title}{Thermal conductivity of crystalline and amorphous
  ices and its implications on amorphization and glassy water}.
\newblock \bibinfo{journal}{Phys. Chem. Chem. Phys.} \bibinfo{volume}{7},
  \bibinfo{pages}{1441--1449}.
\bibitem[{Barr and Showman(2009)}]{barr09}
\bibinfo{author}{Barr, A.}, \bibinfo{author}{Showman, A.},
  \bibinfo{year}{2009}.
\newblock \bibinfo{title}{Heat transfer in europa's icy shell (tucson, az}.
\bibitem[{B^^c4^^9bhounkov^^c3^^a1 et~al.(2021)B^^c4^^9bhounkov^^c3^^a1, Tobie,
  Choblet, Kervazo, Melwani~Daswani, Dumoulin and Vance}]{behounkova20}
\bibinfo{author}{B^^c4^^9bhounkov^^c3^^a1, M.}, \bibinfo{author}{Tobie, G.},
  \bibinfo{author}{Choblet, G.}, \bibinfo{author}{Kervazo, M.},
  \bibinfo{author}{Melwani~Daswani, M.}, \bibinfo{author}{Dumoulin, C.},
  \bibinfo{author}{Vance, S.D.}, \bibinfo{year}{2021}.
\newblock \bibinfo{title}{Tidally induced magmatic pulses on the oceanic floor
  of jupiter's moon europa}.
\newblock \bibinfo{journal}{Geophysical Research Letters} \bibinfo{volume}{48},
  \bibinfo{pages}{e2020GL090077}.
\newblock \URLprefix
  \url{https://agupubs.onlinelibrary.wiley.com/doi/abs/10.1029/2020GL090077},
  \DOIprefix\doi{https://doi.org/10.1029/2020GL090077},
  \href{http://arxiv.org/abs/https://agupubs.onlinelibrary.wiley.com/doi/pdf/10.1029/2020GL090077}{{\tt
  arXiv:https://agupubs.onlinelibrary.wiley.com/doi/pdf/10.1029/2020GL090077}}.
  \bibinfo{note}{e2020GL090077 2020GL090077}.
\bibitem[{Bercovici et~al.(1986)Bercovici, Schubert and Reynolds}]{bercovici86}
\bibinfo{author}{Bercovici, D.}, \bibinfo{author}{Schubert, G.},
  \bibinfo{author}{Reynolds, R.}, \bibinfo{year}{1986}.
\newblock \bibinfo{title}{Phase transitions and convection in icy satellites}.
\newblock \bibinfo{journal}{Geophysical Research Letters} \bibinfo{volume}{13},
  \bibinfo{pages}{448--451}.
\bibitem[{Bercovici et~al.(2022)Bercovici, Elkins-Tanton, O'Rourke and
  Schaefer}]{bercovici22}
\bibinfo{author}{Bercovici, H.L.}, \bibinfo{author}{Elkins-Tanton, L.T.},
  \bibinfo{author}{O'Rourke, J.G.}, \bibinfo{author}{Schaefer, L.},
  \bibinfo{year}{2022}.
\newblock \bibinfo{title}{The effects of bulk composition on planetesimal core
  sulfur content and size}.
\newblock \bibinfo{journal}{Icarus} \bibinfo{volume}{380},
  \bibinfo{pages}{114976}.
\newblock \URLprefix
  \url{https://www.sciencedirect.com/science/article/pii/S0019103522000938},
  \DOIprefix\doi{https://doi.org/10.1016/j.icarus.2022.114976}.
\bibitem[{Bierson and Nimmo(2020)}]{bierson20}
\bibinfo{author}{Bierson, C.J.}, \bibinfo{author}{Nimmo, F.},
  \bibinfo{year}{2020}.
\newblock \bibinfo{title}{Explaining the galilean satellites’ density gradient
  by hydrodynamic escape}.
\newblock \bibinfo{journal}{The Astrophysical Journal Letters}
  \bibinfo{volume}{897}, \bibinfo{pages}{L43}.
\bibitem[{Bland et~al.(2008)Bland, Showman and Tobie}]{bland08}
\bibinfo{author}{Bland, M.T.}, \bibinfo{author}{Showman, A.P.},
  \bibinfo{author}{Tobie, G.}, \bibinfo{year}{2008}.
\newblock \bibinfo{title}{The production of ganymede's magnetic field}.
\newblock \bibinfo{journal}{Icarus} \bibinfo{volume}{198},
  \bibinfo{pages}{384--399}.
\bibitem[{Breuer et~al.(2015)Breuer, Rueckriemen and Spohn}]{breuer15}
\bibinfo{author}{Breuer, D.}, \bibinfo{author}{Rueckriemen, T.},
  \bibinfo{author}{Spohn, T.}, \bibinfo{year}{2015}.
\newblock \bibinfo{title}{Iron snow, crystal floats, and inner^^e2^^80^^93core
  growth: modes of core solidification and implications for dynamos in
  terrestrial planets and moons}.
\newblock \bibinfo{journal}{Progress Earth Planet. Sci.} \bibinfo{volume}{2},
  \bibinfo{pages}{39}.
\bibitem[{Buffett et~al.(1996)Buffett, Huppert, Lister and Woods}]{buffett96}
\bibinfo{author}{Buffett, B.A.}, \bibinfo{author}{Huppert, H.E.},
  \bibinfo{author}{Lister, J.R.}, \bibinfo{author}{Woods, A.W.},
  \bibinfo{year}{1996}.
\newblock \bibinfo{title}{On the thermal evolution of the earth's core}.
\newblock \bibinfo{journal}{Journal of Geophysical Research}
  \bibinfo{volume}{101}, \bibinfo{pages}{7989--8006}.
\bibitem[{Carr et~al.(1998)Carr, Belton, Chapman, Davies, Geissler, Greenberg,
  McEwen, Tufts, Greeley, Sullivan, Head, Pappalardo, Klaasen, Johnson,
  Kaufman, Senske, Moore, Neukum, Schubert, Burns, Thomas and Veverka}]{carr98}
\bibinfo{author}{Carr, M.H.}, \bibinfo{author}{Belton, M.J.S.},
  \bibinfo{author}{Chapman, C.R.}, \bibinfo{author}{Davies, M.E.},
  \bibinfo{author}{Geissler, P.}, \bibinfo{author}{Greenberg, R.},
  \bibinfo{author}{McEwen, A.S.}, \bibinfo{author}{Tufts, B.R.},
  \bibinfo{author}{Greeley, R.}, \bibinfo{author}{Sullivan, R.},
  \bibinfo{author}{Head, J.W.}, \bibinfo{author}{Pappalardo, R.T.},
  \bibinfo{author}{Klaasen, K.P.}, \bibinfo{author}{Johnson, T.V.},
  \bibinfo{author}{Kaufman, J.}, \bibinfo{author}{Senske, D.},
  \bibinfo{author}{Moore, J.}, \bibinfo{author}{Neukum, G.},
  \bibinfo{author}{Schubert, G.}, \bibinfo{author}{Burns, J.A.},
  \bibinfo{author}{Thomas, P.}, \bibinfo{author}{Veverka, J.},
  \bibinfo{year}{1998}.
\newblock \bibinfo{title}{Evidence for a subsurface ocean on europa}.
\newblock \bibinfo{journal}{Nature} \bibinfo{volume}{391},
  \bibinfo{pages}{363--365}.
\bibitem[{Desai(1986)}]{desai86}
\bibinfo{author}{Desai, P.D.}, \bibinfo{year}{1986}.
\newblock \bibinfo{title}{Thermodynamic properties of iron and silicon}.
\newblock \bibinfo{journal}{Journal of Physical and Chemical Reference Data}
  \bibinfo{volume}{15}, \bibinfo{pages}{967--983}.
\bibitem[{Dombard and Sessa(2019)}]{dombard19}
\bibinfo{author}{Dombard, A.J.}, \bibinfo{author}{Sessa, A.M.},
  \bibinfo{year}{2019}.
\newblock \bibinfo{title}{Gravity measurements are key in addressing the
  habitability of a subsurface ocean in jupiter's moon europa}.
\newblock \bibinfo{journal}{Icarus} \bibinfo{volume}{325},
  \bibinfo{pages}{31--38}.
\newblock \URLprefix
  \url{https://www.sciencedirect.com/science/article/pii/S0019103517308527},
  \DOIprefix\doi{https://doi.org/10.1016/j.icarus.2019.02.025}.
\bibitem[{Fei et~al.(1997)Fei, Bertka and Finger}]{fei97}
\bibinfo{author}{Fei, Y.}, \bibinfo{author}{Bertka, C.M.},
  \bibinfo{author}{Finger, L.W.}, \bibinfo{year}{1997}.
\newblock \bibinfo{title}{High-pressure iron-sulfur compound,
  {F}e{$_{3}$}{S}{$_{2}$}, and melting relations in the {F}e-{F}e{S} system}.
\newblock \bibinfo{journal}{Science} \bibinfo{volume}{275},
  \bibinfo{pages}{1621--1623}.
\bibitem[{Goldsby and Kohlstedt(2001)}]{goldsby01}
\bibinfo{author}{Goldsby, D.L.}, \bibinfo{author}{Kohlstedt, D.L.},
  \bibinfo{year}{2001}.
\newblock \bibinfo{title}{Superplastic deformation of ice: Experimental
  observations}.
\newblock \bibinfo{journal}{Journal of Geophysical Research: Solid Earth}
  \bibinfo{volume}{106}, \bibinfo{pages}{11017--11030}.
\bibitem[{{Gomez Casajus} et~al.(2021){Gomez Casajus}, Zannoni, Modenini,
  Tortora, Nimmo, {Van Hoolst}, Buccino and Oudrhiri}]{casajus21}
\bibinfo{author}{{Gomez Casajus}, L.}, \bibinfo{author}{Zannoni, M.},
  \bibinfo{author}{Modenini, D.}, \bibinfo{author}{Tortora, P.},
  \bibinfo{author}{Nimmo, F.}, \bibinfo{author}{{Van Hoolst}, T.},
  \bibinfo{author}{Buccino, D.}, \bibinfo{author}{Oudrhiri, K.},
  \bibinfo{year}{2021}.
\newblock \bibinfo{title}{Updated {E}uropa gravity field and interior structure
  from a reanalysis of {G}alileo tracking data}.
\newblock \bibinfo{journal}{Icarus} \bibinfo{volume}{358},
  \bibinfo{pages}{114187}.
\bibitem[{Grasset and Sotin(1996)}]{grasset96}
\bibinfo{author}{Grasset, O.}, \bibinfo{author}{Sotin, C.},
  \bibinfo{year}{1996}.
\newblock \bibinfo{title}{The cooling rate of a liquid shell in titan's
  interior}.
\newblock \bibinfo{journal}{Icarus} \bibinfo{volume}{123},
  \bibinfo{pages}{101--112}.
\bibitem[{Greenberg et~al.(2000)Greenberg, Geissler, Tufts and
  Hoppa}]{Greenberg00}
\bibinfo{author}{Greenberg, R.}, \bibinfo{author}{Geissler, P.},
  \bibinfo{author}{Tufts, B.R.}, \bibinfo{author}{Hoppa, G.V.},
  \bibinfo{year}{2000}.
\newblock \bibinfo{title}{Habitability of europa's crust: The role of
  tidal-tectonic processes}.
\newblock \bibinfo{journal}{Journal of Geophysical Research: Planets}
  \bibinfo{volume}{105}, \bibinfo{pages}{17551--17562}.
\bibitem[{Hand et~al.(2017)Hand, Murray, J.B.Garvin and Team}]{hand17}
\bibinfo{author}{Hand, K.}, \bibinfo{author}{Murray, A.},
  \bibinfo{author}{J.B.Garvin}, \bibinfo{author}{Team, S.D.},
  \bibinfo{year}{2017}.
\newblock \bibinfo{title}{Report of the europa lander science definition team
  (tech. rep.) pasadena, ca: Nasa} \URLprefix
  \url{https://europa.nasa.gov/resources/58/europa-lander-study-2016-report/}.
\bibitem[{Hauck~II et~al.(2006)Hauck~II, Aurnou and Dombard}]{hauck06}
\bibinfo{author}{Hauck~II, S.A.}, \bibinfo{author}{Aurnou, J.M.},
  \bibinfo{author}{Dombard, A.J.}, \bibinfo{year}{2006}.
\newblock \bibinfo{title}{Sulfur's impact on core evolution and magnetic field
  generation on ganymede}.
\newblock \bibinfo{journal}{Journal of Geophysical Research: Planets}
  \bibinfo{volume}{111}.
\bibitem[{Hill(1962)}]{hill62}
\bibinfo{author}{Hill, M.}, \bibinfo{year}{1962}.
\newblock \bibinfo{title}{The sea. vol. i. physical oceanography: Ideas and
  observations on progress in the study of the sea, new york (interscience
  publishers) 1962. pp ix, 864: Figure. 188s}.
\newblock \bibinfo{journal}{Quarterly Journal of the Royal Meteorological
  Society} \bibinfo{volume}{89}.
\newblock \DOIprefix\doi{10.1002/qj.49708938124}.
\bibitem[{Hirth and Kohlstedt(2004)}]{hirth03}
\bibinfo{author}{Hirth, G.}, \bibinfo{author}{Kohlstedt, D.},
  \bibinfo{year}{2004}.
\newblock \bibinfo{title}{Rheology of the Upper Mantle and the Mantle Wedge: A
  View from the Experimentalists}. \bibinfo{publisher}{American Geophysical
  Union (AGU)}.
\newblock pp. \bibinfo{pages}{83--105}.
\bibitem[{Hobbs(1974)}]{hobbs74}
\bibinfo{author}{Hobbs, P.}, \bibinfo{year}{1974}.
\newblock \bibinfo{title}{Ice physics}.
\newblock \bibinfo{journal}{Oxford University Press, London, UK} .
\bibitem[{Hoppa et~al.(1999)Hoppa, Tufts, Greenberg and Geissler}]{hoppa99}
\bibinfo{author}{Hoppa, G.V.}, \bibinfo{author}{Tufts, B.R.},
  \bibinfo{author}{Greenberg, R.}, \bibinfo{author}{Geissler, P.E.},
  \bibinfo{year}{1999}.
\newblock \bibinfo{title}{Formation of cycloidal features on europa}.
\newblock \bibinfo{journal}{Science} \bibinfo{volume}{285},
  \bibinfo{pages}{1899--1902}.
\bibitem[{Howell(2021)}]{howell21}
\bibinfo{author}{Howell, S.M.}, \bibinfo{year}{2021}.
\newblock \bibinfo{title}{The likely thickness of {E}uropa's icy shell}.
\newblock \bibinfo{journal}{The Planetary Science Journal} \bibinfo{volume}{2},
  \bibinfo{pages}{129}.
\bibitem[{Howell and Pappalardo(2020)}]{howell20}
\bibinfo{author}{Howell, S.M.}, \bibinfo{author}{Pappalardo, R.T.},
  \bibinfo{year}{2020}.
\newblock \bibinfo{title}{{NASA}'s {E}uropa {C}lipper―a mission to a
  potentially habitable ocean world}.
\newblock \bibinfo{journal}{Nature Communications} , \bibinfo{pages}{1311}.
\bibitem[{Hussmann et~al.(2015)Hussmann, Sotin and Lunine}]{hussmann15}
\bibinfo{author}{Hussmann, H.}, \bibinfo{author}{Sotin, C.},
  \bibinfo{author}{Lunine, J.}, \bibinfo{year}{2015}.
\newblock \bibinfo{title}{Interiors and evolution of icy satellites}.
\newblock \bibinfo{journal}{In: G. Schubert (Ed.), Treatise on geophysics 2nd
  edn, Elsevier} , \bibinfo{pages}{605--635}.
\bibitem[{Hussmann and Spohn(2004)}]{hussmann04}
\bibinfo{author}{Hussmann, H.}, \bibinfo{author}{Spohn, T.},
  \bibinfo{year}{2004}.
\newblock \bibinfo{title}{Thermal-orbital evolution of io and europa}.
\newblock \bibinfo{journal}{Icarus} \bibinfo{volume}{171},
  \bibinfo{pages}{391--410}.
\newblock \URLprefix
  \url{https://www.sciencedirect.com/science/article/pii/S0019103504001952},
  \DOIprefix\doi{https://doi.org/10.1016/j.icarus.2004.05.020}.
\bibitem[{Hussmann et~al.(2002)Hussmann, Spohn and Wieczerkowski}]{hussmann02}
\bibinfo{author}{Hussmann, H.}, \bibinfo{author}{Spohn, T.},
  \bibinfo{author}{Wieczerkowski, K.}, \bibinfo{year}{2002}.
\newblock \bibinfo{title}{Thermal equilibrium states of europa's ice shell:
  Implications for internal ocean thickness and surface heat flow}.
\newblock \bibinfo{journal}{Icarus} \bibinfo{volume}{156},
  \bibinfo{pages}{143--151}.
\bibitem[{Kalousov\`{a} et~al.(2018)Kalousov\`{a}, Sotin, Choblet, Tobie and
  Grasset}]{kalousova18}
\bibinfo{author}{Kalousov\`{a}, K.}, \bibinfo{author}{Sotin, C.},
  \bibinfo{author}{Choblet, G.}, \bibinfo{author}{Tobie, G.},
  \bibinfo{author}{Grasset, O.}, \bibinfo{year}{2018}.
\newblock \bibinfo{title}{Two-phase convection in ganymede's high-pressure ice
  layer ―^^c2^^a0implications for its geological evolution}.
\newblock \bibinfo{journal}{Icarus} \bibinfo{volume}{299}, \bibinfo{pages}{133
  -- 147}.
\newblock \URLprefix
  \url{http://www.sciencedirect.com/science/article/pii/S0019103517302403},
  \DOIprefix\doi{https://doi.org/10.1016/j.icarus.2017.07.018}.
\bibitem[{Kamata(2018)}]{kamata18}
\bibinfo{author}{Kamata, S.}, \bibinfo{year}{2018}.
\newblock \bibinfo{title}{One‐dimensional convective thermal evolution
  calculation using a modified mixing length theory: Application to saturnian
  icy satellites}.
\newblock \bibinfo{journal}{J. Geophys. Res.} \bibinfo{volume}{123},
  \bibinfo{pages}{93--112}.
\newblock \DOIprefix\doi{10.1002/2017JE005404}.
\bibitem[{Kamata et~al.(2019)Kamata, Nimmo, Sekine, Kuramoto, Noguchi, Kimura
  and Tani}]{kamata19}
\bibinfo{author}{Kamata, S.}, \bibinfo{author}{Nimmo, F.},
  \bibinfo{author}{Sekine, Y.}, \bibinfo{author}{Kuramoto, K.},
  \bibinfo{author}{Noguchi, N.}, \bibinfo{author}{Kimura, J.},
  \bibinfo{author}{Tani, A.}, \bibinfo{year}{2019}.
\newblock \bibinfo{title}{Pluto's ocean is capped and insulated by gas
  hydrates}.
\newblock \bibinfo{journal}{Nature Geoscience} \bibinfo{volume}{12},
  \bibinfo{pages}{407--410}.
\bibitem[{Karato et~al.(1986)Karato, Peterson and FitzGerald}]{karato86}
\bibinfo{author}{Karato, S.}, \bibinfo{author}{Peterson, M.},
  \bibinfo{author}{FitzGerald, J.}, \bibinfo{year}{1986}.
\newblock \bibinfo{title}{Rheology of synthetic olivine aggregates: {Influence}
  of grain size and water}.
\newblock \bibinfo{journal}{J. Geophys. Res.} \bibinfo{volume}{91},
  \bibinfo{pages}{8151--8176}.
\bibitem[{Kargel(1998)}]{kargel98}
\bibinfo{author}{Kargel, J.S.}, \bibinfo{year}{1998}.
\newblock \bibinfo{title}{Physical Chemistry of Ices in the Outer Solar
  System}. \bibinfo{publisher}{Springer Netherlands},
  \bibinfo{address}{Dordrecht}.
\newblock pp. \bibinfo{pages}{3--32}.
\bibitem[{Kimura and Kamata(2020)}]{kimura2020}
\bibinfo{author}{Kimura, J.}, \bibinfo{author}{Kamata, S.},
  \bibinfo{year}{2020}.
\newblock \bibinfo{title}{Stability of the subsurface ocean of pluto}.
\newblock \bibinfo{journal}{Planetary and Space Science} \bibinfo{volume}{181},
  \bibinfo{pages}{104828}.
\bibitem[{Kimura et~al.(2009)Kimura, Nakagawa and Kurita}]{kimura09}
\bibinfo{author}{Kimura, J.}, \bibinfo{author}{Nakagawa, T.},
  \bibinfo{author}{Kurita, K.}, \bibinfo{year}{2009}.
\newblock \bibinfo{title}{Size and compositional constraints of {Ganymede's}
  metallic core for driving an active dynamo}.
\newblock \bibinfo{journal}{Icarus} \bibinfo{volume}{202},
  \bibinfo{pages}{216--224}.
\bibitem[{Kirk and Stevenson(1987)}]{kirk87}
\bibinfo{author}{Kirk, R.}, \bibinfo{author}{Stevenson, D.},
  \bibinfo{year}{1987}.
\newblock \bibinfo{title}{Thermal evolution of a differentiated ganymede and
  implications for surface features}.
\newblock \bibinfo{journal}{Icarus} \bibinfo{volume}{69}, \bibinfo{pages}{91 --
  134}.
\newblock \URLprefix
  \url{http://www.sciencedirect.com/science/article/pii/0019103587900091},
  \DOIprefix\doi{https://doi.org/10.1016/0019-1035(87)90009-1}.
\bibitem[{Kivelson et~al.(2000)Kivelson, Khurana, Russell, Volwerk, Walker and
  Zimmer}]{kivelson00}
\bibinfo{author}{Kivelson, M.G.}, \bibinfo{author}{Khurana, K.K.},
  \bibinfo{author}{Russell, C.T.}, \bibinfo{author}{Volwerk, M.},
  \bibinfo{author}{Walker, R.J.}, \bibinfo{author}{Zimmer, C.},
  \bibinfo{year}{2000}.
\newblock \bibinfo{title}{Galileo magnetometer measurements: A stronger case
  for a subsurface ocean at europa}.
\newblock \bibinfo{journal}{Science} \bibinfo{volume}{289},
  \bibinfo{pages}{1340--1343}.
\bibitem[{Kuramoto and Matsui(1994)}]{kuramoto94}
\bibinfo{author}{Kuramoto, K.}, \bibinfo{author}{Matsui, T.},
  \bibinfo{year}{1994}.
\newblock \bibinfo{title}{Formation of a hot proto-atmosphere on the accreting
  giant icy satellite: Implications for the origin and evolution of titan,
  ganymede, and callisto}.
\newblock \bibinfo{journal}{Journal of Geophysical Research: Planets}
  \bibinfo{volume}{99}, \bibinfo{pages}{21183--21200}.
\newblock \URLprefix
  \url{https://agupubs.onlinelibrary.wiley.com/doi/abs/10.1029/94JE01864},
  \DOIprefix\doi{https://doi.org/10.1029/94JE01864},
  \href{http://arxiv.org/abs/https://agupubs.onlinelibrary.wiley.com/doi/pdf/10.1029/94JE01864}{{\tt
  arXiv:https://agupubs.onlinelibrary.wiley.com/doi/pdf/10.1029/94JE01864}}.
\bibitem[{Kuskov and Kronrod(2001)}]{kuskov01}
\bibinfo{author}{Kuskov, O.L.}, \bibinfo{author}{Kronrod, V.A.},
  \bibinfo{year}{2001}.
\newblock \bibinfo{title}{Core sizes and internal structure of earth's and
  jupiter's satellites}.
\newblock \bibinfo{journal}{Icarus} \bibinfo{volume}{151},
  \bibinfo{pages}{204--227}.
\newblock \URLprefix
  \url{https://www.sciencedirect.com/science/article/pii/S0019103501966114},
  \DOIprefix\doi{https://doi.org/10.1006/icar.2001.6611}.
\bibitem[{Lamas et~al.(2022)Lamas, Vega and Noya}]{lamas22}
\bibinfo{author}{Lamas, C.P.}, \bibinfo{author}{Vega, C.},
  \bibinfo{author}{Noya, E.G.}, \bibinfo{year}{2022}.
\newblock \bibinfo{title}{Freezing point depression of salt aqueous solutions
  using the madrid-2019 model}.
\newblock \bibinfo{journal}{The Journal of Chemical Physics}
  \bibinfo{volume}{156}, \bibinfo{pages}{134503}.
\bibitem[{Lewis(1972)}]{lewis721}
\bibinfo{author}{Lewis, J.S.}, \bibinfo{year}{1972}.
\newblock \bibinfo{title}{Low temperature condensation from the solar nebula}.
\newblock \bibinfo{journal}{Icarus} \bibinfo{volume}{16},
  \bibinfo{pages}{241--252}.
\bibitem[{Lodders(2003)}]{lodders03}
\bibinfo{author}{Lodders, K.}, \bibinfo{year}{2003}.
\newblock \bibinfo{title}{Solar system abundances and condensation temperatures
  of the elements}.
\newblock \bibinfo{journal}{The Astrophysical Journal} \bibinfo{volume}{591},
  \bibinfo{pages}{1220}.
\bibitem[{Mazarico et~al.(2023)Mazarico, Buccino, Castillo-Rogez, Dombard,
  Genova, Hu{\ss}mann, Kiefer, Lunine, McKinnon, Nimmo, Park, Roberts,
  Srinivasan, Steinbr{\"u}gge, Tortora and Withers}]{mazarico23}
\bibinfo{author}{Mazarico, E.}, \bibinfo{author}{Buccino, D.},
  \bibinfo{author}{Castillo-Rogez, J.}, \bibinfo{author}{Dombard, A.},
  \bibinfo{author}{Genova, A.}, \bibinfo{author}{Hu{\ss}mann, H.},
  \bibinfo{author}{Kiefer, W.}, \bibinfo{author}{Lunine, J.},
  \bibinfo{author}{McKinnon, W.}, \bibinfo{author}{Nimmo, F.},
  \bibinfo{author}{Park, R.}, \bibinfo{author}{Roberts, J.H.},
  \bibinfo{author}{Srinivasan, D.}, \bibinfo{author}{Steinbr{\"u}gge, G.},
  \bibinfo{author}{Tortora, P.}, \bibinfo{author}{Withers, P.},
  \bibinfo{year}{2023}.
\newblock \bibinfo{title}{The europa clipper gravity and radio science
  investigation}.
\newblock \bibinfo{journal}{Space Science Reviews} \bibinfo{volume}{219},
  \bibinfo{pages}{30}.
\newblock \URLprefix \url{https://elib.dlr.de/196823/}.
\bibitem[{Melwani~Daswani et~al.(2021)Melwani~Daswani, Vance, Mayne and
  Glein}]{daswani21}
\bibinfo{author}{Melwani~Daswani, M.}, \bibinfo{author}{Vance, S.D.},
  \bibinfo{author}{Mayne, M.J.}, \bibinfo{author}{Glein, C.R.},
  \bibinfo{year}{2021}.
\newblock \bibinfo{title}{A metamorphic origin for europa's ocean}.
\newblock \bibinfo{journal}{Geophysical Research Letters} \bibinfo{volume}{48},
  \bibinfo{pages}{e2021GL094143}.
\bibitem[{Nimmo(2002)}]{nimmo02}
\bibinfo{author}{Nimmo, F.}, \bibinfo{year}{2002}.
\newblock \bibinfo{title}{{Why does Venus lack a magnetic field?}}
\newblock \bibinfo{journal}{Geology} \bibinfo{volume}{30},
  \bibinfo{pages}{987--990}.
\bibitem[{Nimmo et~al.(2003)Nimmo, Pappalardo and Giese}]{nimmo03}
\bibinfo{author}{Nimmo, F.}, \bibinfo{author}{Pappalardo, R.},
  \bibinfo{author}{Giese, B.}, \bibinfo{year}{2003}.
\newblock \bibinfo{title}{On the origins of band topography, europa}.
\newblock \bibinfo{journal}{Icarus} \bibinfo{volume}{166},
  \bibinfo{pages}{21--32}.
\bibitem[{Ojakangas and Stevenson(1989)}]{ojakangas89}
\bibinfo{author}{Ojakangas, G.}, \bibinfo{author}{Stevenson, D.},
  \bibinfo{year}{1989}.
\newblock \bibinfo{title}{Thermal state of an ice shell on {Europa}}.
\newblock \bibinfo{journal}{Icarus} \bibinfo{volume}{81},
  \bibinfo{pages}{220--241}.
\bibitem[{Pappalardo et~al.(1999)Pappalardo, Belton, Breneman, Carr, Chapman,
  Collins, Denk, Fagents, Geissler, Giese, Greeley, Greenberg, Head,
  Helfenstein, Hoppa, Kadel, Klaasen, Klemaszewski, Magee, McEwen, Moore,
  Moore, Neukum, Phillips, Prockter, Schubert, Senske, Sullivan, Tufts, Turtle,
  Wagner and Williams}]{pappalardo99}
\bibinfo{author}{Pappalardo, R.T.}, \bibinfo{author}{Belton, M.J.S.},
  \bibinfo{author}{Breneman, H.H.}, \bibinfo{author}{Carr, M.H.},
  \bibinfo{author}{Chapman, C.R.}, \bibinfo{author}{Collins, G.C.},
  \bibinfo{author}{Denk, T.}, \bibinfo{author}{Fagents, S.},
  \bibinfo{author}{Geissler, P.E.}, \bibinfo{author}{Giese, B.},
  \bibinfo{author}{Greeley, R.}, \bibinfo{author}{Greenberg, R.},
  \bibinfo{author}{Head, J.W.}, \bibinfo{author}{Helfenstein, P.},
  \bibinfo{author}{Hoppa, G.}, \bibinfo{author}{Kadel, S.D.},
  \bibinfo{author}{Klaasen, K.P.}, \bibinfo{author}{Klemaszewski, J.E.},
  \bibinfo{author}{Magee, K.}, \bibinfo{author}{McEwen, A.S.},
  \bibinfo{author}{Moore, J.M.}, \bibinfo{author}{Moore, W.B.},
  \bibinfo{author}{Neukum, G.}, \bibinfo{author}{Phillips, C.B.},
  \bibinfo{author}{Prockter, L.M.}, \bibinfo{author}{Schubert, G.},
  \bibinfo{author}{Senske, D.A.}, \bibinfo{author}{Sullivan, R.J.},
  \bibinfo{author}{Tufts, B.R.}, \bibinfo{author}{Turtle, E.P.},
  \bibinfo{author}{Wagner, R.}, \bibinfo{author}{Williams, K.K.},
  \bibinfo{year}{1999}.
\newblock \bibinfo{title}{Does europa have a subsurface ocean? evaluation of
  the geological evidence}.
\newblock \bibinfo{journal}{Journal of Geophysical Research: Planets}
  \bibinfo{volume}{104}, \bibinfo{pages}{24015--24055}.
\bibitem[{Patankar(1980)}]{patankar80}
\bibinfo{author}{Patankar, S.}, \bibinfo{year}{1980}.
\newblock \bibinfo{title}{Numerical heat transfer and fluid flow}.
\newblock \bibinfo{journal}{Hemisphere Publishing Corporation, New York.} .
\bibitem[{Pommier(2018)}]{pommier18}
\bibinfo{author}{Pommier, A.}, \bibinfo{year}{2018}.
\newblock \bibinfo{title}{Influence of sulfur on the electrical resistivity of
  a crystallizing core in small terrestrial bodies}.
\newblock \bibinfo{journal}{Earth and Planetary Science Letters}
  \bibinfo{volume}{496}, \bibinfo{pages}{37--46}.
\bibitem[{Rathbun et~al.(1998)Rathbun, Musser~Jr. and Squyres}]{Rathbun98}
\bibinfo{author}{Rathbun, J.A.}, \bibinfo{author}{Musser~Jr., G.S.},
  \bibinfo{author}{Squyres, S.W.}, \bibinfo{year}{1998}.
\newblock \bibinfo{title}{Ice diapirs on europa: Implications for liquid
  water}.
\newblock \bibinfo{journal}{Geophysical Research Letters} \bibinfo{volume}{25},
  \bibinfo{pages}{4157--4160}.
\bibitem[{R^^c3^^bcckriemen et~al.(2018)R^^c3^^bcckriemen, Breuer and
  Spohn}]{ruckriemen18}
\bibinfo{author}{R^^c3^^bcckriemen, T.}, \bibinfo{author}{Breuer, D.},
  \bibinfo{author}{Spohn, T.}, \bibinfo{year}{2018}.
\newblock \bibinfo{title}{Top-down freezing in a fe^^e2^^80^^93fes core and
  ganymede’s present-day magnetic field}.
\newblock \bibinfo{journal}{Icarus} \bibinfo{volume}{307},
  \bibinfo{pages}{172--196}.
\bibitem[{Roberts et~al.(2023)Roberts, McKinnon, Elder, Tobie, Biersteker,
  Young, Park, Steinbrugge, Nimmo, Howell, Castillo-Rogez, Cable, Abrahams,
  Bland, Chivers, Cochrane, Dombard, Ernst, Genova, Gerekos, Glein, Harris,
  Hay, Hayne, Hedman, Hussmann, Jia, Khurana, Kiefer, Kirk, Kivelson, Lawrence,
  Leonard, Lunine, Mazarico, McCord, McEwen, Paty, Quick, Raymond, Retherford,
  Roth, Rymer, Saur, Scanlan, Schroeder, Senske, Shao, Soderlund, Spiers,
  Styczinski, Tortora, Vance, Villarreal, Weiss, Westlake, Withers,
  Wolfenbarger, Buratti, Korth, Pappalardo and Group}]{roberts23}
\bibinfo{author}{Roberts, J.H.}, \bibinfo{author}{McKinnon, W.B.},
  \bibinfo{author}{Elder, C.M.}, \bibinfo{author}{Tobie, G.},
  \bibinfo{author}{Biersteker, J.B.}, \bibinfo{author}{Young, D.},
  \bibinfo{author}{Park, R.S.}, \bibinfo{author}{Steinbrugge, G.},
  \bibinfo{author}{Nimmo, F.}, \bibinfo{author}{Howell, S.M.},
  \bibinfo{author}{Castillo-Rogez, J.C.}, \bibinfo{author}{Cable, M.L.},
  \bibinfo{author}{Abrahams, J.N.}, \bibinfo{author}{Bland, M.T.},
  \bibinfo{author}{Chivers, C.}, \bibinfo{author}{Cochrane, C.J.},
  \bibinfo{author}{Dombard, A.J.}, \bibinfo{author}{Ernst, C.},
  \bibinfo{author}{Genova, A.}, \bibinfo{author}{Gerekos, C.},
  \bibinfo{author}{Glein, C.}, \bibinfo{author}{Harris, C.D.},
  \bibinfo{author}{Hay, H.C.F.C.}, \bibinfo{author}{Hayne, P.O.},
  \bibinfo{author}{Hedman, M.}, \bibinfo{author}{Hussmann, H.},
  \bibinfo{author}{Jia, X.}, \bibinfo{author}{Khurana, K.},
  \bibinfo{author}{Kiefer, W.S.}, \bibinfo{author}{Kirk, R.},
  \bibinfo{author}{Kivelson, M.}, \bibinfo{author}{Lawrence, J.},
  \bibinfo{author}{Leonard, E.J.}, \bibinfo{author}{Lunine, J.I.},
  \bibinfo{author}{Mazarico, E.}, \bibinfo{author}{McCord, T.B.},
  \bibinfo{author}{McEwen, A.}, \bibinfo{author}{Paty, C.},
  \bibinfo{author}{Quick, L.C.}, \bibinfo{author}{Raymond, C.A.},
  \bibinfo{author}{Retherford, K.D.}, \bibinfo{author}{Roth, L.},
  \bibinfo{author}{Rymer, A.}, \bibinfo{author}{Saur, J.},
  \bibinfo{author}{Scanlan, K.}, \bibinfo{author}{Schroeder, D.M.},
  \bibinfo{author}{Senske, D.A.}, \bibinfo{author}{Shao, W.},
  \bibinfo{author}{Soderlund, K.}, \bibinfo{author}{Spiers, E.},
  \bibinfo{author}{Styczinski, M.J.}, \bibinfo{author}{Tortora, P.},
  \bibinfo{author}{Vance, S.D.}, \bibinfo{author}{Villarreal, M.N.},
  \bibinfo{author}{Weiss, B.P.}, \bibinfo{author}{Westlake, J.H.},
  \bibinfo{author}{Withers, P.}, \bibinfo{author}{Wolfenbarger, N.},
  \bibinfo{author}{Buratti, B.}, \bibinfo{author}{Korth, H.},
  \bibinfo{author}{Pappalardo, R.T.}, \bibinfo{author}{Group, T.I.T.W.},
  \bibinfo{year}{2023}.
\newblock \bibinfo{title}{Exploring the interior of europa with the europa
  clipper}.
\newblock \bibinfo{journal}{Space Science Reviews} \bibinfo{volume}{219},
  \bibinfo{pages}{46}.
\newblock \URLprefix \url{https://doi.org/10.1007/s11214-023-00990-y},
  \DOIprefix\doi{10.1007/s11214-023-00990-y}.
\bibitem[{Sasaki and Nakazawa()}]{sasaki86}
\bibinfo{author}{Sasaki, S.}, \bibinfo{author}{Nakazawa, K.}, .
\newblock \bibinfo{title}{Metal-silicate fractionation in the growing earth:
  Energy source for the terrestrial magma ocean}.
\newblock \bibinfo{journal}{Journal of Geophysical Research: Solid Earth}
  \bibinfo{volume}{91}, \bibinfo{pages}{9231--9238}.
\bibitem[{Schenk(2002)}]{schenk02}
\bibinfo{author}{Schenk, P.}, \bibinfo{year}{2002}.
\newblock \bibinfo{title}{Thickness constraints on the icy shells of the
  galilean satellites from a comparison of crater shapes}.
\newblock \bibinfo{journal}{Nature} , \bibinfo{pages}{419--421}.
\bibitem[{Schubert et~al.(1986)Schubert, Spohn and Reynolds}]{schubert86}
\bibinfo{author}{Schubert, G.}, \bibinfo{author}{Spohn, T.},
  \bibinfo{author}{Reynolds, R.}, \bibinfo{year}{1986}.
\newblock \bibinfo{title}{Thermal histories, and internal structures of the
  moons of the solar system}.
\newblock \bibinfo{journal}{In: Burns, J. A., Matthews, M. S. (Eds.),
  Satellites. Univ. of Arizona Press, Tucson} , \bibinfo{pages}{224--292}.
\bibitem[{Senshu et~al.(2002)Senshu, Kuramoto and Matsui}]{senshu02}
\bibinfo{author}{Senshu, H.}, \bibinfo{author}{Kuramoto, K.},
  \bibinfo{author}{Matsui, T.}, \bibinfo{year}{2002}.
\newblock \bibinfo{title}{Thermal evolution of a growing mars}.
\newblock \bibinfo{journal}{Journal of Geophysical Research: Planets}
  \bibinfo{volume}{107}, \bibinfo{pages}{1--1--1--13}.
\bibitem[{Showman et~al.(1997)Showman, Stevenson and Malhotra}]{showman97}
\bibinfo{author}{Showman, A.P.}, \bibinfo{author}{Stevenson, D.J.},
  \bibinfo{author}{Malhotra, R.}, \bibinfo{year}{1997}.
\newblock \bibinfo{title}{Coupled orbital and thermal evolution of {Ganymede}}.
\newblock \bibinfo{journal}{Icarus} \bibinfo{volume}{129},
  \bibinfo{pages}{367--383}.
\bibitem[{Sohl et~al.(2002)Sohl, Spohn, Breuer and Nagel}]{sohl02}
\bibinfo{author}{Sohl, F.}, \bibinfo{author}{Spohn, T.},
  \bibinfo{author}{Breuer, D.}, \bibinfo{author}{Nagel, K.},
  \bibinfo{year}{2002}.
\newblock \bibinfo{title}{Implications from galileo observations on the
  interior structure and chemistry of the galilean satellites}.
\newblock \bibinfo{journal}{Icarus} \bibinfo{volume}{157},
  \bibinfo{pages}{104--119}.
\bibitem[{Sotin et~al.(1998)Sotin, Grasset and Beauchesne}]{sotin98}
\bibinfo{author}{Sotin, C.}, \bibinfo{author}{Grasset, O.},
  \bibinfo{author}{Beauchesne, S.}, \bibinfo{year}{1998}.
\newblock \bibinfo{title}{Thermodynamic properties of high pressure ices:
  implications for the dynamics and internal structure of large icy
  satellites}.
\newblock \bibinfo{journal}{In: B. Schmitt~et~al. (Eds.), Solar System Ices.
  Kluwer Academic Press, Dordrecht} , \bibinfo{pages}{79--96}.
\bibitem[{Sotin et~al.(2009)Sotin, Tobie, Wahr and McKinnon}]{sotin09}
\bibinfo{author}{Sotin, C.}, \bibinfo{author}{Tobie, G.},
  \bibinfo{author}{Wahr, J.}, \bibinfo{author}{McKinnon, W.B.},
  \bibinfo{year}{2009}.
\newblock \bibinfo{title}{Tides and tidal heating on Europa}.
  \bibinfo{publisher}{The University of Arizona Press},
  \bibinfo{address}{Tucson}.
\newblock pp. \bibinfo{pages}{85--118}.
\bibitem[{St^^c3^^a4hler et~al.(2018)St^^c3^^a4hler, Panning, Vance, Lorenz,
  van^^c2^^a0Driel, Nissen-Meyer and Kedar}]{stahler18}
\bibinfo{author}{St^^c3^^a4hler, S.C.}, \bibinfo{author}{Panning, M.P.},
  \bibinfo{author}{Vance, S.D.}, \bibinfo{author}{Lorenz, R.D.},
  \bibinfo{author}{van^^c2^^a0Driel, M.}, \bibinfo{author}{Nissen-Meyer, T.},
  \bibinfo{author}{Kedar, S.}, \bibinfo{year}{2018}.
\newblock \bibinfo{title}{Seismic wave propagation in icy ocean worlds}.
\newblock \bibinfo{journal}{Journal of Geophysical Research: Planets}
  \bibinfo{volume}{123}, \bibinfo{pages}{206--232}.
\bibitem[{Stevenson(2003)}]{stevenson03}
\bibinfo{author}{Stevenson, D.J.}, \bibinfo{year}{2003}.
\newblock \bibinfo{title}{Planetary magnetic fields}.
\newblock \bibinfo{journal}{Earth and Planetary Science Letters}
  \bibinfo{volume}{208}, \bibinfo{pages}{1--11}.
\bibitem[{Tobie et~al.(2003)Tobie, Choblet and Sotin}]{tobie03}
\bibinfo{author}{Tobie, G.}, \bibinfo{author}{Choblet, G.},
  \bibinfo{author}{Sotin, C.}, \bibinfo{year}{2003}.
\newblock \bibinfo{title}{Tidally heated convection: Constraints on europa's
  ice shell thickness}.
\newblock \bibinfo{journal}{Journal of Geophysical Research: Planets}
  \bibinfo{volume}{108}.
\bibitem[{Travis et~al.(2012)Travis, Palguta and Schubert}]{travis12}
\bibinfo{author}{Travis, B.}, \bibinfo{author}{Palguta, J.},
  \bibinfo{author}{Schubert, G.}, \bibinfo{year}{2012}.
\newblock \bibinfo{title}{A whole-moon thermal history model of europa: Impact
  of hydrothermal circulation and salt transport}.
\newblock \bibinfo{journal}{Icarus} \bibinfo{volume}{218},
  \bibinfo{pages}{1006--1019}.
\bibitem[{Trinh et~al.(2023)Trinh, Bierson and O'Rourke}]{trinh23}
\bibinfo{author}{Trinh, K.T.}, \bibinfo{author}{Bierson, C.J.},
  \bibinfo{author}{O'Rourke, J.G.}, \bibinfo{year}{2023}.
\newblock \bibinfo{title}{Slow evolution of {E}uropa's interior: metamorphic
  ocean origin, delayed metallic core formation, and limited seafloor
  volcanism}.
\newblock \bibinfo{journal}{Science Advances} \bibinfo{volume}{9},
  \bibinfo{pages}{eadf3955}.
\newblock \URLprefix
  \url{https://www.science.org/doi/abs/10.1126/sciadv.adf3955},
  \DOIprefix\doi{10.1126/sciadv.adf3955},
  \href{http://arxiv.org/abs/https://www.science.org/doi/pdf/10.1126/sciadv.adf3955}{{\tt
  arXiv:https://www.science.org/doi/pdf/10.1126/sciadv.adf3955}}.
\bibitem[{Trumbo et~al.(2022)Trumbo, Becker, Brown, Denman, Molyneux, Hendrix,
  Retherford, Roth and Alday}]{trumbo22}
\bibinfo{author}{Trumbo, S.K.}, \bibinfo{author}{Becker, T.M.},
  \bibinfo{author}{Brown, M.E.}, \bibinfo{author}{Denman, W.T.P.},
  \bibinfo{author}{Molyneux, P.}, \bibinfo{author}{Hendrix, A.},
  \bibinfo{author}{Retherford, K.D.}, \bibinfo{author}{Roth, L.},
  \bibinfo{author}{Alday, J.}, \bibinfo{year}{2022}.
\newblock \bibinfo{title}{A new uv spectral feature on europa: Confirmation of
  nacl in leading-hemisphere chaos terrain}.
\newblock \bibinfo{journal}{The Planetary Science Journal} \bibinfo{volume}{3},
  \bibinfo{pages}{27}.
\bibitem[{Trumbo et~al.(2019)Trumbo, Brown and Hand}]{trumbo19}
\bibinfo{author}{Trumbo, S.K.}, \bibinfo{author}{Brown, M.E.},
  \bibinfo{author}{Hand, K.P.}, \bibinfo{year}{2019}.
\newblock \bibinfo{title}{Sodium chloride on the surface of europa}.
\newblock \bibinfo{journal}{Science Advances} \bibinfo{volume}{5},
  \bibinfo{pages}{eaaw7123}.
\bibitem[{Turtle and Pierazzo(2001)}]{turtle01}
\bibinfo{author}{Turtle, E.P.}, \bibinfo{author}{Pierazzo, E.},
  \bibinfo{year}{2001}.
\newblock \bibinfo{title}{Thickness of a europan ice shell from impact crater
  simulations}.
\newblock \bibinfo{journal}{Science} \bibinfo{volume}{294},
  \bibinfo{pages}{1326--1328}.
\bibitem[{Vance et~al.(2023)Vance, Craft, Shock, Schmidt, Lunine, Hand,
  McKinnon, Spiers, Chivers, Lawrence, Wolfenbarger, Leonard, Robinson,
  Styczinski, Persaud, Steinbrugge, Zolotov, Quick, Scully, Becker,
  Howell, Clark, Dombard, Glein, Mousis, Sephton, Castillo-Rogez, Nimmo,
  McEwen, Gudipati, Jun, Jia, Postberg, Soderlund and Elder}]{vance23}
\bibinfo{author}{Vance, S.}, \bibinfo{author}{Craft, K.},
  \bibinfo{author}{Shock, E.}, \bibinfo{author}{Schmidt, B.},
  \bibinfo{author}{Lunine, J.}, \bibinfo{author}{Hand, K.},
  \bibinfo{author}{McKinnon, W.}, \bibinfo{author}{Spiers, E.},
  \bibinfo{author}{Chivers, C.}, \bibinfo{author}{Lawrence, J.},
  \bibinfo{author}{Wolfenbarger, N.}, \bibinfo{author}{Leonard, E.},
  \bibinfo{author}{Robinson, K.}, \bibinfo{author}{Styczinski, M.},
  \bibinfo{author}{Persaud, D.}, \bibinfo{author}{Steinbrugge, G.},
  \bibinfo{author}{Zolotov, M.}, \bibinfo{author}{Quick, L.},
  \bibinfo{author}{Scully, J.}, \bibinfo{author}{Becker, T.},
  \bibinfo{author}{Howell, S.}, \bibinfo{author}{Clark, R.},
  \bibinfo{author}{Dombard, A.}, \bibinfo{author}{Glein, C.},
  \bibinfo{author}{Mousis, O.}, \bibinfo{author}{Sephton, M.},
  \bibinfo{author}{Castillo-Rogez, J.}, \bibinfo{author}{Nimmo, F.},
  \bibinfo{author}{McEwen, A.}, \bibinfo{author}{Gudipati, M.},
  \bibinfo{author}{Jun, I.}, \bibinfo{author}{Jia, X.},
  \bibinfo{author}{Postberg, F.}, \bibinfo{author}{Soderlund, K.},
  \bibinfo{author}{Elder, C.}, \bibinfo{year}{2023}.
\newblock \bibinfo{title}{Investigating europa's habitability with the europa
  clipper}.
\newblock \bibinfo{journal}{Space Science Reviews} \bibinfo{volume}{219}.
\newblock \URLprefix \url{http://dx.doi.org/10.1007/s11214-023-01025-2},
  \DOIprefix\doi{10.1007/s11214-023-01025-2}.
\bibitem[{Vilella et~al.(2020)Vilella, Choblet, Tsao and Deschamps}]{vilella20}
\bibinfo{author}{Vilella, K.}, \bibinfo{author}{Choblet, G.},
  \bibinfo{author}{Tsao, W.E.}, \bibinfo{author}{Deschamps, F.},
  \bibinfo{year}{2020}.
\newblock \bibinfo{title}{Tidally heated convection and the occurrence of
  melting in icy satellites: Application to europa}.
\newblock \bibinfo{journal}{Journal of Geophysical Research: Planets}
  \bibinfo{volume}{125}, \bibinfo{pages}{e2019JE006248}.
\newblock \bibinfo{note}{E2019JE006248 10.1029/2019JE006248}.
\bibitem[{Wagner et~al.(2011)Wagner, Sohl, Hussmann, Grott and
  Rauer}]{wagner11}
\bibinfo{author}{Wagner, F.}, \bibinfo{author}{Sohl, F.},
  \bibinfo{author}{Hussmann, H.}, \bibinfo{author}{Grott, M.},
  \bibinfo{author}{Rauer, H.}, \bibinfo{year}{2011}.
\newblock \bibinfo{title}{Interior structure models of solid exoplanets using
  material laws in the infinite pressure limit}.
\newblock \bibinfo{journal}{Icarus} \bibinfo{volume}{214}, \bibinfo{pages}{366
  -- 376}.
\bibitem[{Wasson and Kallemeyn(1988)}]{Wasson535}
\bibinfo{author}{Wasson, J.}, \bibinfo{author}{Kallemeyn, G.},
  \bibinfo{year}{1988}.
\newblock \bibinfo{title}{Compositions of chondrites}.
\newblock \bibinfo{journal}{Philosophical Transactions of the Royal Society of
  London A: Mathematical, Physical and Engineering Sciences}
  \bibinfo{volume}{325}, \bibinfo{pages}{535--544}.
\bibitem[{Williams(2009)}]{williams09}
\bibinfo{author}{Williams, Q.}, \bibinfo{year}{2009}.
\newblock \bibinfo{title}{Bottom-up versus top-down solidification of the cores
  of small solar system bodies: Constraints on paradoxical cores}.
\newblock \bibinfo{journal}{Earth and Planetary Science Letters}
  \bibinfo{volume}{284}, \bibinfo{pages}{564--569}.
\bibitem[{Zhan and Schubert(2012)}]{zhan12}
\bibinfo{author}{Zhan, X.}, \bibinfo{author}{Schubert, G.},
  \bibinfo{year}{2012}.
\newblock \bibinfo{title}{Powering ganymede's dynamo}.
\newblock \bibinfo{journal}{Journal of Geophysical Research: Planets}
  \bibinfo{volume}{117}.

\end{thebibliography}

\end{document}